\documentclass[letterpaper,amsmath,amssymb,aps,
prd,preprint,nofootinbib,superscriptaddress]{revtex4}

\usepackage{color}

\usepackage[T1]{fontenc} 
\usepackage{lmodern}
\usepackage{url}
\usepackage{graphicx,subfigure}

\usepackage{amsmath,amssymb}
\usepackage{amsfonts}
\usepackage{graphicx, rotating}
\usepackage{epstopdf}
\usepackage{epsfig}
\usepackage{latexsym}
\usepackage{bm}
\usepackage[utf8]{inputenc}

\usepackage{float}
\usepackage{ulem}

\usepackage{hyperref}
\hypersetup{ setpagesize=false, bookmarks=true,bookmarksnumbered=true, colorlinks=true,linkcolor=blue, citecolor=red, linktocpage=true, urlcolor=blue, hypertexnames=true}
\usepackage{natbib}

\newcommand{\lagr}{\mathcal{L}}
\newcommand{\anomalies}{(g-2)_{e,\mu}}
\newcommand{\bea}{\begin{eqnarray}}
	\newcommand{\eea}{\end{eqnarray}}

\newcommand{\be}{\begin{equation}}
	\newcommand{\ee}{\end{equation}}

\def\lsim{\raise0.3ex\hbox{$\;<$\kern-0.75em\raise-1.1ex\hbox{$\sim\;$}}}
\def\gsim{\raise0.3ex\hbox{$\;>$\kern-0.75em\raise-1.1ex\hbox{$\sim\;$}}}

\DeclareUnicodeCharacter{2212}{-}
\DeclareUnicodeCharacter{2032}{'}

\begin{document} 
	\preprint{ULB-TH/21-01}
	\title{ \boldmath Solving the electron and muon $g-2$ anomalies in $Z'$ models}

	\author{Arushi Bodas}
	\email[]{arushib@terpmail.umd.edu}
	\affiliation{Maryland Center for Fundamental Physics, University of Maryland, College Park, MD 20742}
	
	\author{Rupert Coy}
	\email[]{rupert.coy@ulb.be}
	\affiliation{Service de Physique Th\'eorique, Universit\'e Libre de Bruxelles,
		Boulevard du Triomphe, CP225, 1050 Brussels, Belgium}
	
	\author{Simon J.D. King}
	\email[]{sjd.king@soton.ac.uk}
	\affiliation{\small School of Physics and Astronomy, University of Southampton,
		Southampton, SO17 1BJ, United Kingdom}
	%
	
	%

	
	\begin{abstract}
		
		
		We consider simultaneous explanations of the electron and muon $g-2$ anomalies through a single $Z'$ of a $U(1)'$ extension to the Standard Model (SM). 
		We first perform a model-independent analysis of the viable flavour-dependent $Z'$ couplings to leptons, which are subject to various strict experimental constraints.
		We show that only a narrow region of parameter space with an MeV-scale $Z'$ can account for the two anomalies.
		Following the conclusions of this analysis, we then explore the ability of different classes of $Z'$ models to realise these couplings, including the SM$+U(1)'$, the $N$-Higgs Doublet Model$+U(1)'$, and a Froggatt-Nielsen style scenario.
		In each case, the necessary combination of couplings cannot be obtained, owing to additional relations between the $Z'$ couplings to charged leptons and neutrinos induced by the gauge structure, and to the stringency of neutrino scattering bounds.
		Hence, we conclude that no $U(1)'$ extension can resolve both anomalies unless other new fields are also introduced. 
		While most of our study assumes the Caesium $(g-2)_e$ measurement, our findings in fact also hold in the case of the Rubidium measurement, despite the tension between the two.

	\end{abstract}

	\maketitle

	\tableofcontents
	

	\section{Introduction}
	\label{sec:Intro}
	
	The excellent agreement between the Standard Model (SM) and experimental observations makes the persisting anomalies all the more interesting. 
	One long-standing discrepancy between theory and experiment is that of the anomalous magnetic dipole moment of the muon, $a_{\mu} \equiv (g-2)_{\mu}/2$, which has recently been updated to a $4.2\sigma$ tension with the SM \cite{Bennett:2006fi, Aoyama:2020ynm,Abi:2021gix}\footnote{We note that the significance of this anomaly has been questioned by a lattice QCD calculation of the leading-order hadronic vacuum polarisation contribution to $a_\mu^\text{SM}$ \cite{Borsanyi:2020mff}.
	},
	\begin{equation}
		\Delta a_{\mu} \equiv a_\mu^\text{exp} - a_\mu^\text{SM} = (2.51 \pm 0.59) \times 10^{-9} \, .
		\label{muonanomaly}
	\end{equation} 
	Further data from the ongoing Muon g-2 experiment at Fermilab is expected to reduce the uncertainty by a factor of four  \cite{Grange:2015fou}, and the future J-PARC experiment forecasts similar precision \cite{Abe:2019thb}, both of which should clarify the status of this disagreement. 
	To add to the puzzle, an anomaly emerged in the electron sector due to a) an improved measurement of fine-structure constant, $\alpha_{\rm em}$, using Caesium atoms \cite{Parker:2018vye}, from which the value of $(g-2)_e$ may be extracted, and b) an updated theoretical calculation \cite{Aoyama:2017uqe}. 
	This yielded a discrepancy in the electron anomalous magnetic moment of
	\begin{equation}
		\Delta a_e^{\text{Cs}} \equiv a_e^\text{exp~(\text{Cs})} - a_e^\text{SM} = (-8.7 \pm 3.6) \times 10^{-13} \, ,
		\label{electronanomaly}
	\end{equation}
	which constitutes a $2.4 \sigma$ tension with the SM \cite{Davoudiasl:2018fbb}. 
	Notably, this has the opposite sign to the muon anomaly, Eq. \eqref{muonanomaly}. 
	Recently, however, a new measurement of the fine-structure constant using Rubidium atoms gave \cite{Morel:2020dww} 
	\begin{equation}
		\Delta a_e^{\text{Rb}} \equiv a_e^\text{exp~(\text{Rb})} - a_e^\text{SM} = (4.8 \pm 3.0) \times 10^{-13} \, .
		\label{electronanomalyRub}
	\end{equation}
	This is a milder anomaly, the discrepancy between experiment and SM being only 1.6$\sigma$, and it is in the same direction as the muon anomaly. 
	Remarkably, the Caesium and Rubidium measurements of $\alpha_{\rm em}$ disagree by more than $5\sigma$, therefore it is difficult to obtain a consistent picture of $a_e^\text{exp}$.

	Given this uncertain status quo, in this paper we choose to focus predominantly on the earlier Caesium result, Eq.~\eqref{electronanomaly}, and only discuss the Rubidium result in section \ref{sec:Rubidium} (which, however, is the first $Z'$ analysis of this new experimental situation, to the best of our knowledge).
	The presence of dual anomalies in the electron and muon sectors motivates an exploration of new physics models that could simultaneously explain both. 
	Moreover, the relative size and sign of these anomalies poses an interesting theoretical challenge.

	Let us consider these issues. 
	Firstly, the opposite signs of $\Delta a_\mu$ and $\Delta a_e^{\text{Cs}}$ (from now on we will drop the superscript) immediately excludes all new physics models whose contribution to the magnetic dipole moment of charged leptons has a fixed sign. 
	The dark photon \cite{Holdom:1985ag}, for instance, generates $\Delta a_{e,\mu} > 0$, and therefore cannot satisfy the dual anomalies. 
	Secondly, the contribution from flavour-universal new physics to $(g-2)$ is generally expected to be proportional to the mass or mass squared of the lepton (see e.g. \cite{Freitas:2014pua,Dorsner:2016wpm}), whereas from Eqs.~\eqref{muonanomaly} and \eqref{electronanomaly} we find
	\begin{align}
		\frac{m_e^2}{m_\mu^2} \ll \left| \frac{\Delta a_e}{\Delta a_{\mu}} \right| \sim 3.5 \times 10^{-4} \ll \frac{m_e}{m_\mu}  \, .
	\end{align}
	These considerations, along with numerous low scale constraints discussed below, lead to significant model-building obstacles. 
	So far, various attempts have been made to explain the anomalies, with different solutions relying on the introduction of new scalars, SUSY, leptoquarks, vector-like fermions, or other BSM mechanisms, see e.g. \cite{Crivellin:2018qmi,Endo:2019bcj,Badziak:2019gaf,Hiller:2019mou,Bauer:2019gfk,Endo:2020mev,Giudice:2012ms,Davoudiasl:2018fbb,Liu:2018xkx,Dutta:2018fge,Gardner:2019mcl,Cornella:2019uxs,Dutta:2020scq,Yang:2020bmh,Crivellin:2019mvj,Bigaran:2020jil,Dorsner:2020aaz,Botella:2020xzf,Jana:2020pxx,Han:2018znu,Abdullah:2019ofw,CarcamoHernandez:2020pxw,Haba:2020gkr,Calibbi:2020emz,Arbelaez:2020rbq,Chen:2020jvl,Hati:2020fzp,Jana:2020joi,Chen:2020tfr,Chun:2020uzw,Li:2020dbg,Banerjee:2020zvi,Hernandez:2021tii,Darme:2020sjf}. 
	In this paper, we study a rather unexplored possibility that a (light) $Z'$ boson with flavour-dependent lepton couplings accounts for both anomalies.

	A new gauge boson of a $U'(1)$ symmetry is a well-motivated candidate for many BSM models. 
	It has long been considered a possible explanation of the $(g-2)_\mu$ anomaly \cite{Pospelov:2008zw} (see also e.g. \cite{Davoudiasl:2014kua,1511.07447,1712.09360,1901.09552,1906.11297}), thus it seems important to investigate if a $U(1)'$ extension of the SM can at the same time also resolve the $(g-2)_e$ anomaly.
	One immediate advantage of $Z'$ models is that it is possible to generate positive or negative contributions to the magnetic moment simply by adjusting the relative size of its vector and axial couplings to fermions, as will be shown below.

	We focus on the $Z'$ in mass range $m_e < m_{Z'} < m_\mu $, which is a natural consequence of various experimental bounds (more on this in Sections \ref{subsec:Anomalies} and \ref{sec:constraints}). 
	A $Z'$ in the MeV mass range has been of interest (see e.g. \cite{Feng:2016ysn,Kozaczuk:2016nma,Lindner:2018kjo,Bauer:2018onh,DelleRose:2018eic,Smolkovic:2019jow}) due to hints of a new $17$ MeV boson to explain anomalies in nuclear transitions observed by the Atomki collaboration, both in Beryllium \cite{Krasznahorkay:2015iga}, and more recently Helium \cite{Krasznahorkay:2019lyl}. Models with MeV-scale $Z'$ also have the capacity to generate $\Delta N_\text{eff} \simeq 0.2$ in the early Universe \cite{Escudero:2019gzq}, thereby somewhat ameliorating the Hubble tension \cite{Bernal:2016gxb}. 
	The question then is whether the scenario survives the wealth of sensitive experiments, in particular for $m_{Z'} \sim \mathcal{O}(\text{MeV})$. To answer this, we first perform a model-independent analysis to identify regions in the parameter space of $Z'$ models that can successfully explain both the $(g-2)$ anomalies. This to our knowledge is the first study of this scenario in such a general and model-independent way, although a specific $Z'$ model was previously studied in the context of the dual $(g-2)$ anomalies and found not to work \cite{CarcamoHernandez:2019ydc}. 
	Note that we are focusing on the minimal scenario where the additional contribution to the anomalous magnetic moments comes solely from the $Z'$, which is different from some of the other models studied in literature that include a $Z'$ plus other new fields (e.q. \cite{Hati:2020fzp,Banerjee:2020zvi,CarcamoHernandez:2019ydc}).
	The conclusions from our model-independent analysis serve as a powerful tool in checking the viability of various specific $Z'$ models, and we hope that it will be useful for more complex model-building.


	The layout is as follows: 
	Section \ref{sec:Formalism} introduces our conventions for the effective $Z'$ couplings and potential origins of these couplings.
	We study experimental constraints on these couplings in Section \ref{sec:constraints}, summarising our findings in Figs. \ref{fig:e_constraints} and \ref{fig:mu_constraints}. 
	In light of the array of experiments probing light vector bosons in the near future, we discuss the discovery potential of such a $Z'$ in Section \ref{sec:future}. Equipped with the model-independent analysis, in Section \ref{sec:models} we consider several models and the challenges they face. 
	We demonstrate that some of the simplest and most common classes of $U(1)'$ extensions of the SM cannot explain the two anomalies simultaneously.
	Finally, in Section \ref{sec:Rubidium} we address the Rubidium $(g-2)_e$ anomaly and study the capacity of a $Z'$ model to explain it in conjunction with the $(g-2)_\mu$ anomaly.

	
	\section{Formalism of a light $Z'$}
	\label{sec:Formalism}

	\subsection{Effective $Z'$ couplings}
	In the most general framework, a new $Z'$ with family-dependent charged lepton couplings leads to flavour violation. However, in this paper we assume that the charged lepton Yukawa matrix and the matrix of charged lepton $Z'$ couplings 
	are simultaneously diagonalisable and therefore the $Z'$ has only lepton-flavour conserving couplings. Various flavour models predict such scenarios (see, for instance, \cite{Criado:2019tzk}) and in this way we avoid stringent limits on flavour-violation, such as from $\mu \rightarrow e \gamma$ \cite{TheMEG:2016wtm}.
	Flavour-conserving couplings of fermions to the $Z'$ can be described through $\mathcal{L} = - Z'_\mu J^\mu_{Z'}$, with gauge current,
	\begin{equation}
		J_{Z'} ^\mu = \sum _f \bar{\psi}_f \gamma ^\mu ( C_{L f}P_L + C_{R f} P_R) \psi _f \, .
	\end{equation}
	Rewriting the charged lepton interactions in terms of vector and axial couplings, $C_{V(A)~f} = (C_{R f}\pm C_{L f})/2$, gives 
	\begin{equation}
		\mathcal{L} \supset - \sum \limits_{\alpha = e,\mu,\tau} \left[ \overline{\ell_\alpha} \gamma^\mu \left( C_{V\alpha} + C_{A \alpha} \gamma_5 \right) \ell_\alpha + C_{\nu \alpha} \overline{\nu_\alpha} \gamma^\mu P_L \nu_\alpha \right] Z'_\mu ~.
		\label{effectivecouplings}
	\end{equation}
	It is typically a simple exercise to derive these effective couplings for a given model. 
	For now we assume that the different effective couplings are unrelated. 
	In models with no extra fermions, there are three different contributions to the couplings of SM fermions to the $Z'$ arising from a $U(1)'$ gauge group. 
	These are:
	\begin{itemize} 
		\item Charge assignment of the fermion under the $U(1)'$ ({\textit{flavour dependent}}).
		\item Gauge-Kinetic Mixing (GKM) arising from 
		the Lagrangian term $\mathcal{L} ^\textrm{GKM} = -\frac{\varepsilon}{2} B'_{\mu \nu} X'^{\mu \nu}$, where $B'_{\mu \nu}$ and $X'_{\mu \nu}$ are the field strength tensors of hypercharge and $U(1)'$, respectively ({\textit{flavour universal}}).
		\item $Z-Z'$ mass mixing, which is generated if the SM Higgs sector is charged under the $U(1)'$ ({\textit{flavour universal}}).
	\end{itemize}
	The combination of these three contributions can generate variety of vector and axial couplings.
	As explained in the introduction, in this work we are concerned with exploring the possibility that 
	a single $Z'$ accounts for the $(g-2)_{e,\mu}$ discrepancies. 
	We will firstly survey the parameter space in a model-independent way in terms of the effective lepton-$Z'$ couplings defined in Eq. \eqref{effectivecouplings}. 
	The conclusions from this analysis are then used in Sections \ref{sec:models} and \ref{sec:Rubidium} to study whether these couplings can be realised in a few specific classes of $Z'$ models.

	\begin{figure}[t!]
		\centering
		\includegraphics[scale=0.2]{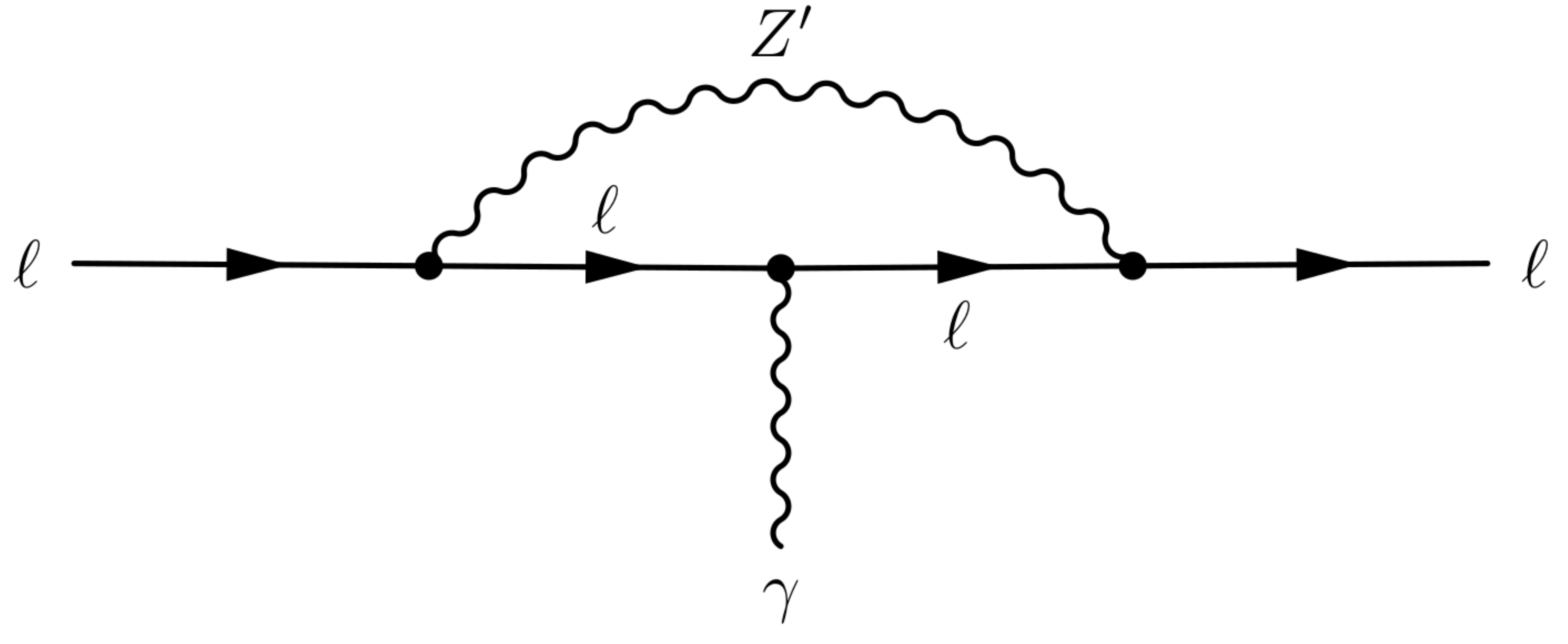} 
		\caption{\small The one-loop $Z'$ contribution to the anomalous magnetic moment of a charged lepton.}
		\label{fig:FeynmanDiag}
	\end{figure}

	\subsection{Contribution to the charged lepton anomalous magnetic moment}
	\label{subsec:Anomalies}
	The $Z'$ modifies the magnetic moment of a charged lepton via the one-loop diagram in Fig.~\ref{fig:FeynmanDiag}. 
	In the notation of Eq.~\eqref{effectivecouplings}, the contribution for a charged lepton of flavour $\alpha$ is \cite{Leveille:1977rc}
	\begin{equation}
		\Delta a_\alpha = \frac{m_\alpha^2}{4\pi^2 m_{Z'}^2} \left( C_{V\alpha}^2 \int_0^1 \frac{x^2 (1-x)}{1 - x + x^2 m_\alpha^2/m_{Z'}^2} dx - C_{A\alpha}^2 \int_0^1 \frac{x (1-x) (4 - x) + 2 x^3 m_\alpha^2/ m_{Z'}^2}{1 - x + x^2 m_\alpha^2/m_{Z'}^2} dx \right) \, .
		\label{gm2sfull}
	\end{equation}
	In the limits $m_\alpha \ll m_{Z'}$ and $m_\alpha \gg m_{Z'}$, this simplifies to 
	\begin{equation}
		\Delta a_\alpha \simeq \begin{cases}
			m_\alpha^2 \left( C_{V\alpha}^2 - 5 C_{A\alpha}^2 \right)/(12\pi^2 m_{Z'}^2)~, & m_\alpha \ll m_{Z'} \, , \\
			(m_{Z'}^2 C_{V\alpha}^2 - 2m_\alpha^2 C_{A\alpha}^2)/(8\pi^2 m_{Z'}^2) ~, & m_\alpha \gg m_{Z'} \, .
		\end{cases}
		\label{gm2s}
	\end{equation}
	We see that the way to achieve correct signs for the contributions to muon and electron anomalies ($\Delta a_e < 0$ and $\Delta a_\mu >0$) is with a non-zero axial coupling for the electron ($C_{Ae}$) and vector coupling for the muon ($C_{V\mu}$).

	
	We remark that it is impossible to satisfy both the anomalies simultaneously if we demand flavour universality, i.e. $C_{Ve} = C_{V \mu}$ and $C_{Ae} = C_{A \mu}$. 
	This is straightforward to see from  Eq.~\eqref{gm2s} when $m_{Z'} < m_e$ or $m_{Z'} > m_\mu$. 
	For the remaining case of $m_e < m_{Z'} < m_{\mu}$, solving the anomalies demands $|C_{Ae}| \gtrsim 0.45 \, |C_{Ve}|$ and $|C_{V\mu}| \gg |C_{A\mu}|$, which is inconsistent with flavour universality.\footnote{It is interesting to note that if the anomalies had the opposite sign, i.e. had the experimental data required $\Delta a_e > 0$ and $\Delta a_\mu < 0$, then $C_{Ve} = C_{V \mu}$ and $C_{Ae} = C_{A \mu}$ could have given a viable solution. 
		Thus, neither the different sign nor the unusual ratio of the anomalies necessarily implies that flavour non-universal physics must be present.}
	This is precisely why we consider models with flavour-dependent $Z'$ couplings in this paper.

	We may now make some broad arguments about preferred $m_{Z'}$ values.
	In the case of a light $Z'$ with $m_{Z'}\ll m_e$, even the smallest effective couplings required to explain the anomalies, accomplished by setting $C_{Ve}=C_{A \mu}=0$, lead to orders of magnitude between $C_{V \mu}$ and $C_{A e}$, which could only be accounted for by either an orders of magnitude difference in their charges under the $U(1)'$ or a very fine-tuned cancellation of the flavour-dependent part of $C_{Ae}$ against the flavour-universal contribution.
	We will see in Section \ref{sec:constraints} that such a light $Z'$ with couplings sufficiently large that it satisfies the anomalies is in any case excluded by cosmological constraints. 
	Therefore, we will focus on $m_{Z'} > m_e$. 
	
	For the heavy regime, i.e. $m_{Z'} \gg m_\mu$, two arguments follow. Firstly, considering the muon sector, in the region $2m_{\mu} < m_{Z'} \leq 10 \textrm{ GeV}$, the new vector boson is excluded by BaBar from its decay into two muons \cite{BaBar:2016sci}, while for $5$ GeV $\leq m_{Z'} \leq 70$ GeV it is similarly excluded by CMS \cite{CMS:2018yxg}. Turning to the electron sector, we note that for $m_{Z'} \gtrsim 10 $ GeV, the axial coupling to electrons required to satisfy the anomaly in electron sector is $|C_{Ae}| \gtrsim 0.1$. 
	With such large couplings to electrons, any GeV-scale object would have likely showed a signal at previous colliders, such as SLAC RF linac, LEP and LHC runs. 
	A heavy $Z'$ solution to the anomalies therefore seems improbable given these considerations.
	Finally, in the intermediate range, $m_e < m_{Z'}< m_\mu$, the values of $C_{Ae}$ and $C_{V\mu}$ required to explain the two anomalies are of a similar order of magnitude, which lends this mass range to potentially more natural, i.e. less fine-tuned, solutions and so we will focus on this regime in the remainder of this paper.
	

	\section{Model-independent analysis of Constraints on $Z'$ couplings}
	\label{sec:constraints}
	The effective couplings introduced in Eq.~\eqref{effectivecouplings} are subject to a wide variety of constraints, which we shall now discuss. 
	In general, the $Z'$ could couple to all SM fermions, and indeed there are some rather stringent bounds on $Z'$ couplings to quarks. 
	However, we will focus on $Z'$ interactions with electrons and muons, those being the critical ones for the explanation of the $\anomalies$ anomalies. 
	Since lepton doublets contain both charged leptons and neutrinos, non-zero effective couplings to charged leptons generally imply effective couplings to neutrinos, which have their own experimental constraints. 
	This will be borne out in the example models considered in Section \ref{sec:models}.\footnote{The dark photon is a notable counter-example, with interactions solely generated through gauge-kinetic mixing, where $C_{V\alpha} \neq 0$ while $C_{A\alpha} = C_{\nu \alpha} = 0$. 
		However, the dark photon does not successfully explain the $\anomalies$ anomalies because, as is easily seen from Eq.~\eqref{gm2sfull}, $C_{Ae} = 0$ implies $\Delta a_e \geq 0$. }

	For a given explicit model, there may be many additional constraints. 
	These can arise in several different ways. 
	Firstly, as mentioned just above, the $Z'$ may also couple to the tau or to quarks. 
	Bounds on $Z'$ couplings to light quarks are discussed for instance in \cite{Feng:2016ysn,Kozaczuk:2016nma,Bauer:2018onh,DelleRose:2018eic,Smolkovic:2019jow}. 
	Secondly, $Z-Z'$ mixing leads to a shift in $Z$ boson couplings, which have been very precisely measured at LEP \cite{ALEPH:2005ab}, as well as other electroweak-scale parameters. 
	
	While there may be such model-dependent bounds, the goal of this section is to study the viability or otherwise of a $Z'$ solution to the two anomalies based on leptonic $Z'$ couplings alone. 
	The plethora of experimental constraints are described below, with our results summarised in Figs.~\ref{fig:e_constraints} and \ref{fig:mu_constraints}.

	\subsection{Couplings to electrons}
	\label{electroncouplings}
	We first outline the most important limits on the effective couplings of the $Z'$ to electrons ($C_{Ve}$, $C_{Ae}$) and electron neutrinos ($C_{\nu e}$).

	\subsubsection{Cosmological and astrophysical bounds}
	MeV-scale states with even very small interactions with electrons or neutrinos (effective couplings as tiny as $|C| \sim 10^{-9}$) can remain in thermal contact with the SM plasma during Big Bang Nucleosynthesis (BBN) and thereby significantly alter early universe cosmology. 
	Bounds on the masses of electrophilic and neutrinophilic vector bosons from various cosmological probes were calculated in \cite{Sabti:2019mhn}. Combining BBN and Planck data, they found at 95.4\% C.L. that an electrophilic $Z'$, i.e. $\sqrt{C_{Ve}^2 + C_{Ae}^2} \gg |C_{\nu e}|$, is constrained to have a mass of at least 9.8 MeV. 
	From Eqs.~\eqref{electronanomaly} and \eqref{gm2s}, we see that for $m_{Z'} \gtrsim $ MeV, the effective electron-$Z'$ coupling should be $|C_{Ae}| > 10^{-6}$, so the BBN bounds do apply here. 
	The limit is slightly weakened for larger $|C_{\nu e}|$, therefore we take $m_{Z'} \geq 9.8$ MeV as a conservative lower bound on our $Z'$ mass.\footnote{This bound can in principle be avoided by sufficiently light $m_{Z'}$. 
		When $m_{Z'} \lesssim 100$ eV, the $(g-2)_e$ anomaly can be explained with $|C_{Ve}|, |C_{Ae}|, |C_{\nu e}| < 10^{-9}$.
		However, we will see in Sec.~\ref{muoncouplings} that such a light $Z'$ solution to the $(g-2)_\mu$ discrepancy is ruled out by similar cosmological considerations.
	}

	The $Z'$ also affects various aspects of stellar evolution. 
	The most critical of these for a MeV-scale $Z'$ is white dwarf cooling \cite{Dreiner:2013tja}. 
	The $Z'$ mediates an additional source of cooling, via $e^+ e^- \to Z' \to \nu \bar{\nu}$. 
	Since the $Z'$ mass under consideration is much larger than white dwarf temperatures, $T_{WD} \sim 5$ keV, this can be treated as an effective four-fermion interaction at the scale $T_{WD}$ with the $Z'$ integrated out. 
	Motivated by the good agreement between predictions and observations of white dwarf cooling, the benchmark set by \cite{Dreiner:2013tja} is that new sources of cooling should not exceed SM ones. 
	We therefore impose 
	\begin{equation}
		\frac{\sqrt{(C_{Ve}^2 + C_{Ae}^2) (C^2_{\nu e}+C^2_{\nu \mu}+C^2_{\nu \tau})}}{m_{Z'}^2} \leq G_F \, ,
		\label{wdconstraint}
	\end{equation}
	as an approximate bound. 
	When plotting this constraint in Fig.~\ref{fig:e_constraints}, we assume that only $C_{Ve}$, $C_{Ae}$ and $C_{\nu e}$ are non-zero.

	Finally, we note that a $Z'$ which couples to neutrinos can also be an additional source of energy loss for supernovae, if it is able to escape the supernova core. 
	We followed the formalism in Appendix B of \cite{Escudero:2019gzq} and enforced that the additional energy loss due to the $Z'$ is no greater than the energy loss in the SM during the first ten seconds of the supernova explosion.
	However, for a roughly MeV-scale $Z'$ this observation only constrains a band of effective couplings $10^{-12} \lesssim C_{\nu \alpha} \lesssim 10^{-7}$, which is much too small to be relevant for the anomalies.

	\subsubsection{Collider and beam dump bounds}
	
	A stringent limit on $Z'$ interactions with electrons comes from the BaBar experiment, which searched for a dark photon, $A'$, via $e^+ e^- \to \gamma A'$ with $A' \to e^+ e^-$. 
	The results are reported in \cite{Lees:2014xha} and probe masses from 20 MeV up to 10.2 GeV. 
	The bound on $\varepsilon$, the kinetic mixing parameter in the dark photon model arising from the gauge-kinetic term $\lagr ^\textrm{GKM} \supset -\frac{\varepsilon}{2}B'_{\mu \nu} X'^{\mu \nu}$ can be converted into a limit on $\sqrt{C_{Ve}^2 + C_{Ae}^2}$. 
	We neglect the statistical fluctuations in the BaBar bound (cf. Fig. 4 of \cite{Lees:2014xha}), opting conservatively to extrapolate from the most constraining points of the 90\% confidence exclusion region and obtain our bound by interpolating between these. 
	This constraint becomes mildly stronger with $Z'$ mass, with for instance
	\begin{equation}
		\sqrt{(C_{Ve}^2 + C_{Ae}^2) \text{BR}(Z' \to e^+ e^-)} \lesssim 3 (6) \times 10^{-4} \, ,
		\label{babarconstraint}
	\end{equation}
	for $m_{Z'} \gtrsim 40(20)$ MeV.
	For $m_{Z'} < 2m_\mu$, the $Z'$ is sufficiently light that it decays only to electrons and neutrinos. 
	We will see that the couplings to electrons should generically be much larger than couplings to neutrinos, thus $\text{BR}(Z' \to e^+ e^-) \approx 1$.

	The BaBar result alone rules out a vast region of parameter space. 
	The smallest axial $Z'-e$ coupling required to satisfy the $(g-2)_e$ discrepancy is given by $|C_{Ae}| \simeq 9\times 10^{-6} (m_{Z'}/\text{MeV})$, as can be seen from Eq.~\eqref{gm2s} by setting $C_{Ve} = 0$. 
	Then the BaBar bound $|C_{Ae}| \lesssim 3\times 10^{-4}$ for $m_{Z'} \gtrsim 40$ MeV  rules out all solutions (notwithstanding some statistical fluctuations) with a $Z'$ heavier than 40 MeV up to the largest mass probed by the experiment, 10.2 GeV. 
	This limit only strengthens for $C_{Ve} \neq 0$, since a larger $C_{Ae}$ is then required to explain $(g-2)_e$, while at the same time $C_{Ae}$ is more constrained because BaBar bounds the combination $\sqrt{C_{Ve}^2 + C_{Ae}^2}$.

	The KLOE experiment also constrains the $Z'$ coupling to electrons \cite{Anastasi:2015qla}. 
	Although generally weaker than BaBar's limit, its exclusion region covers additional parameter space since the experiment probes masses as small as 5 MeV. For these low masses, the bound is around 
	\begin{equation}
		\sqrt{(C_{Ve}^2 + C_{Ae}^2) \text{BR}(Z' \to e^+ e^-)} \lesssim 6 \times 10^{-4} \, .
		\label{kloeconstraint}
	\end{equation}

	Beam dump experiments probe the $Z'$ couplings to electrons, since the $Z'$ may be produced and detected via $e^- + Z \to e^- +  Z'[\to e^+ e^-]$, see e.g. \cite{Bjorken:2009mm}. 
	The produced $Z'$s should therefore decay in the dump before they reach the detector. 
	The best bound comes from NA64 \cite{Banerjee:2019hmi}, which sets limits on a $Z'$ with masses between 1 MeV and 24 MeV.

	A further stringent bound on the parameter space comes from the precise measurement of parity-violating M\o ller scattering at SLAC \cite{Anthony:2005pm}. 
	For $Z'$ masses below around 100 MeV, the bound is independent of $m_{Z'}$ and yields \cite{Kahn:2016vjr}
	\begin{equation}
		|C_{Ve} C_{Ae}| \lesssim 10^{-8} \, .
		\label{moller}
	\end{equation}
	As indicated above, a tiny $C_{Ve}$ is ideal for explaining the $(g-2)_e$ anomaly while avoiding collider constraints with as small a value of $\sqrt{C_{Ve}^2 + C_{Ae}^2}$ as possible. 
	Taking $C_{Ve}$ close to zero is clearly also an efficient way to evade this M\o ller scattering limit.

	\subsubsection{Neutrino scattering bounds}

	Very strong restrictions on the effective couplings come from measurements of neutrino-charged lepton scattering \cite{Lindner:2018kjo}. 
	There have been many experiments testing neutrino interactions. Here we study the most relevant ones: TEXONO \cite{Deniz:2009mu} Borexino \cite{Bellini:2011rx}, and CHARM-II \cite{Vilain:1993kd}. 
	These experiments are known to be among the most constraining in general (see e.g. \cite{Feng:2016ysn,Lindner:2018kjo,Bauer:2018onh,Smolkovic:2019jow}), they cover a range of energies and different neutrino flavours. 
	Let us first consider TEXONO. The typical energy transfer in a scattering event is $\sim \sqrt{m_e T}$, where $3 \textrm{~MeV} \leq T \leq 8 $ MeV is the electron recoil energy. With $m_{Z'} \gtrsim 10$ MeV (as enforced by the limits from cosmology), we may safely make the assumption $m_{Z'} \gg \sqrt{m_e T}$. 
	The correction to the SM cross-section of anti-neutrino scattering is then
	\begin{align}
		\frac{\sigma(\overline{\nu_e}  e^- \to \overline{\nu_e}  e^-)}{\sigma(\overline{\nu_e} e^- \to \overline{\nu_e}  e^-)^\text{SM}} \simeq  1 & + \left(2.07 C_{Ve} + 1.39 C_{Ae} \right) 10^{11} C_{\nu e} \left(\frac{\text{MeV}}{m_{Z'}} \right)^2 \notag \\
		& + \left( 1.37 C_{Ve}^2 + 2.62 C_{Ve} C_{Ae} + 1.64 C_{Ae}^2 \right) \left( 10^{11} C_{\nu e} \right)^2 \left( \frac{\text{MeV}}{m_{Z'}} \right)^4 ~,
		\label{texono}
	\end{align} 
	following Ref. \cite{Lindner:2018kjo}. 
	Comparing this with the TEXONO measurement, $\sigma(\overline{\nu_e}  e^- \to \overline{\nu_e}  e^-)^\text{exp} = (1.08 \pm 0.26) \times \sigma(\overline{\nu_e}  e^- \to \overline{\nu_e}  e^-)^\text{SM}$ \cite{Deniz:2009mu} puts extremely stringent bounds on the $Z'$ effective couplings.

	Borexino measures the scattering of solar neutrinos. 
	The electron neutrino survival probability is measured as $(51 \pm 7)\%$, while the experiment cannot distinguish muon and tau neutrinos. 
	For simplicity, we therefore assume that 50\% of the scattered neutrinos are electron neutrinos, with 25\% each of muon and tau neutrinos.\footnote{This assumption has negligible bearing on our main results since $C_{\nu \mu}$ is specifically probed by CHARM-II, as outlined below, and we are not concerned with $C_{\nu \tau}$.} 
	Then the scattering rate induced by the $Z'$ is 
	\begin{align}
		&\frac{\sigma(\nu e^- \to \nu  e^-)}{\sigma(\nu e^- \to \nu  e^-)^\text{SM}} \simeq 1 + 10^{10} \left(\frac{\text{MeV}}{m_{Z'}} \right)^2 \Big[ C_{Ve} \left( 6.86 C_{\nu e} - 1.16 C_{\nu\mu} - 1.16 C_{\nu\tau} \right) \notag \\
		&+ C_{Ae} \left( - 8.27 C_{\nu e} + 2.22 C_{\nu\mu} + 2.22 C_{\nu\tau} \right) \Big] \notag \\
		&+ 10^{21} \left(\frac{\text{MeV}}{m_{Z'}} \right)^4 \left( 1.38 C_{Ae}^2 + 0.81 C_{Ve }^2 - 1.38 C_{Ve} C_{Ae} \right) \left( 2 C_{\nu e}^2 + C_{\nu\mu}^2 + C_{\nu\tau }^2 \right)   ~. 
		\label{borexino}
	\end{align}
	The cross-section including new physics should not deviate from the SM cross-section by more than about 10$\%$ \cite{Bellini:2011rx,Harnik:2012ni}, and this restriction sets a strong limit on the parameter space. 
	Note from Eqs.~\eqref{texono} and \eqref{borexino} that $C_{Ve}$ and $C_{Ae}$ can both be large as long as the $C_{\nu e,\mu,\tau}$ are sufficiently small. 

	\begin{figure}[hbt!]
		\subcapraggedrighttrue
		\subfigure[\footnotesize{\sl $C_{\nu e}=0$, $|C_{Ve}|$ fixed such that $a_{e}$ is $1\sigma$ below the experimental value}.]{\includegraphics[height=0.35\textwidth, width=0.46\textwidth]{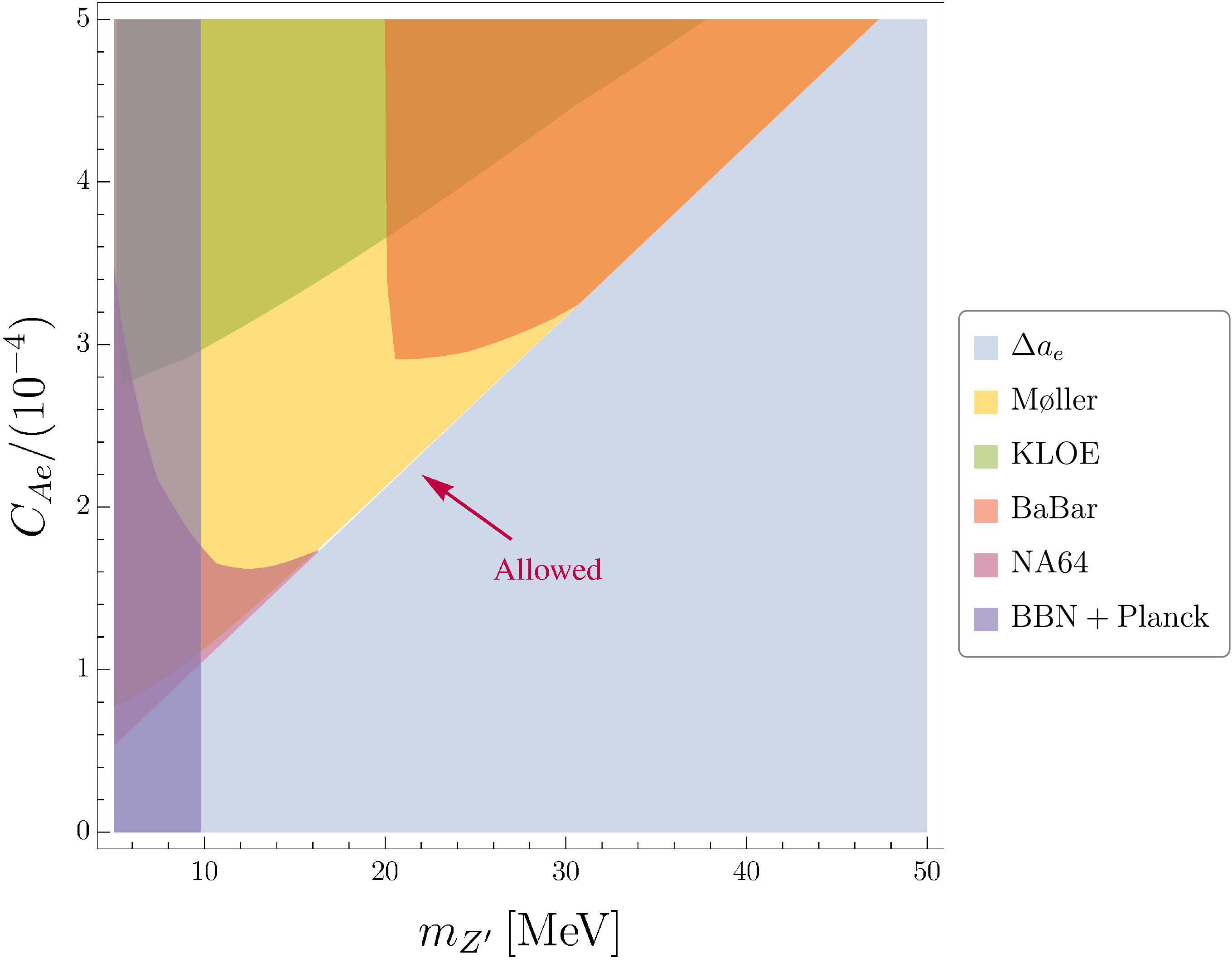}} \quad
		\subfigure[\footnotesize{\sl $C_{\nu e}=0$, $|C_{Ve}|$ fixed such that $a_{e}$ is $1\sigma$ above the experimental value}.]{\includegraphics[height=0.35\textwidth,width=0.46\textwidth]{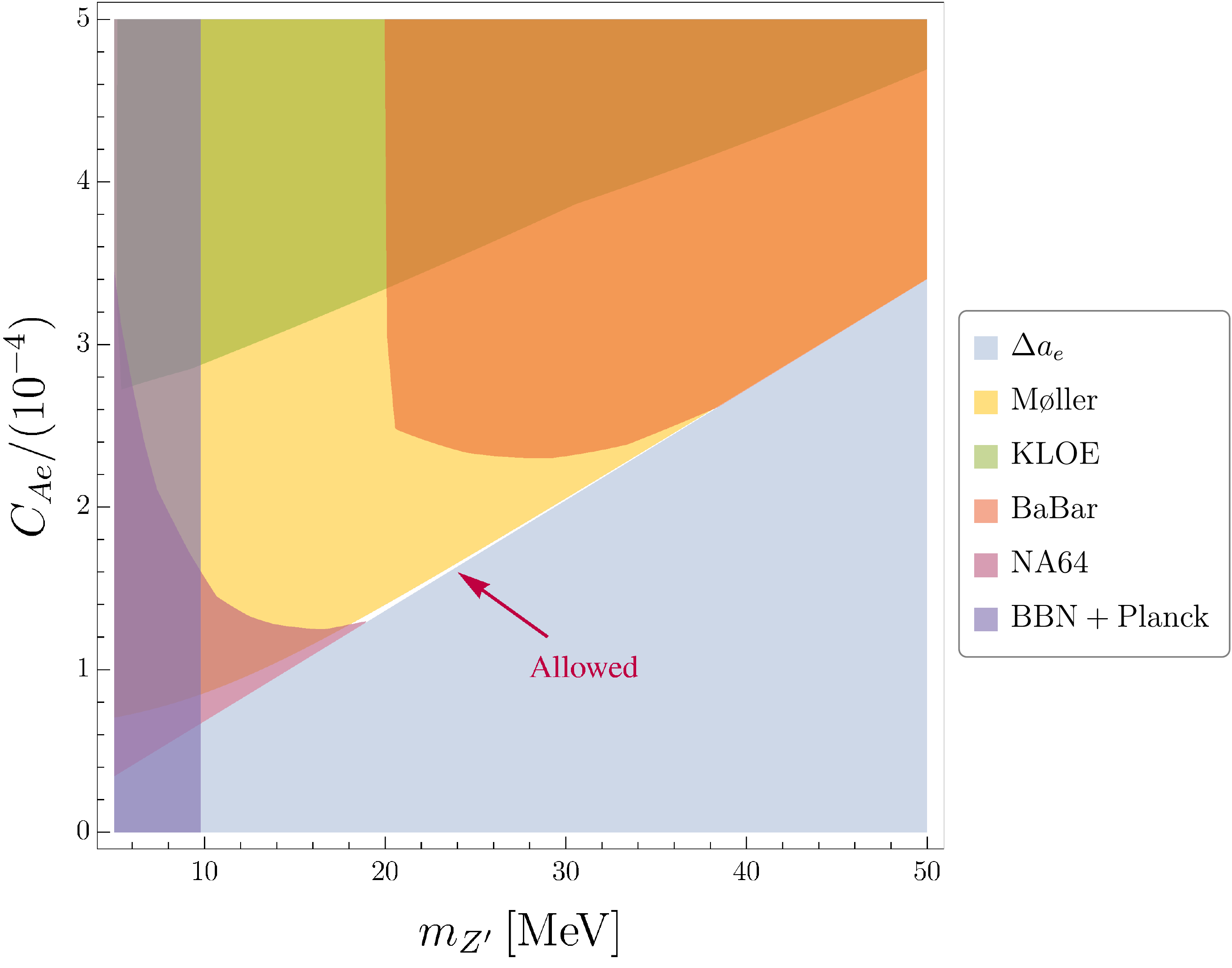}} 
		\subfigure[\footnotesize{\sl $C_{Ve}=0$, $C_{Ae}>0$ fixed such that $a_{e}$ is $1\sigma$ below the experimental value}.]{\includegraphics[height=0.35\textwidth,width=0.46\textwidth]{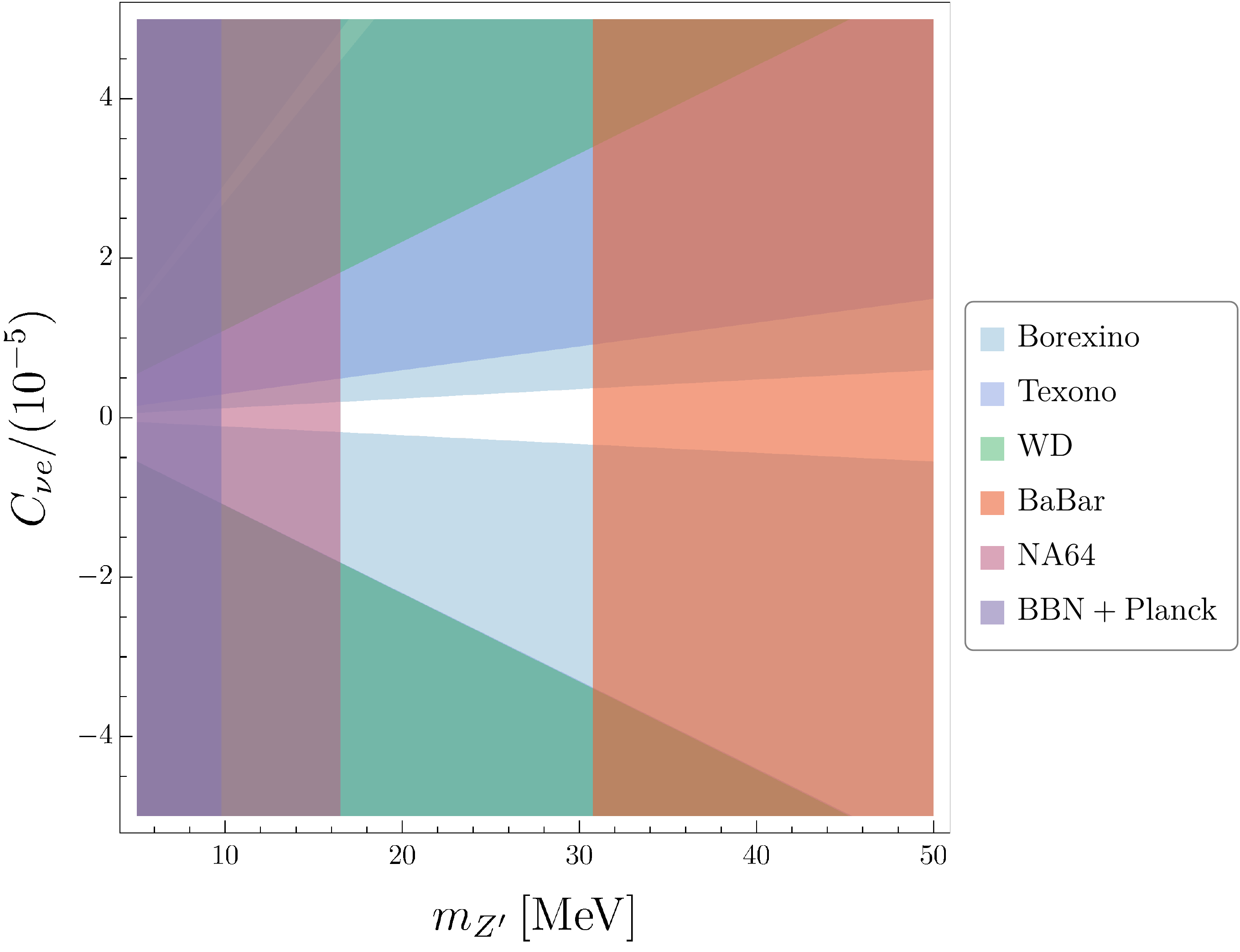}} \quad
		\subfigure[\footnotesize{\sl $C_{Ve}=0$, $C_{Ae}>0$ fixed such that $a_{e}$ is $1\sigma$ above the experimental value}.]{\includegraphics[height=0.35\textwidth,width=0.46\textwidth]{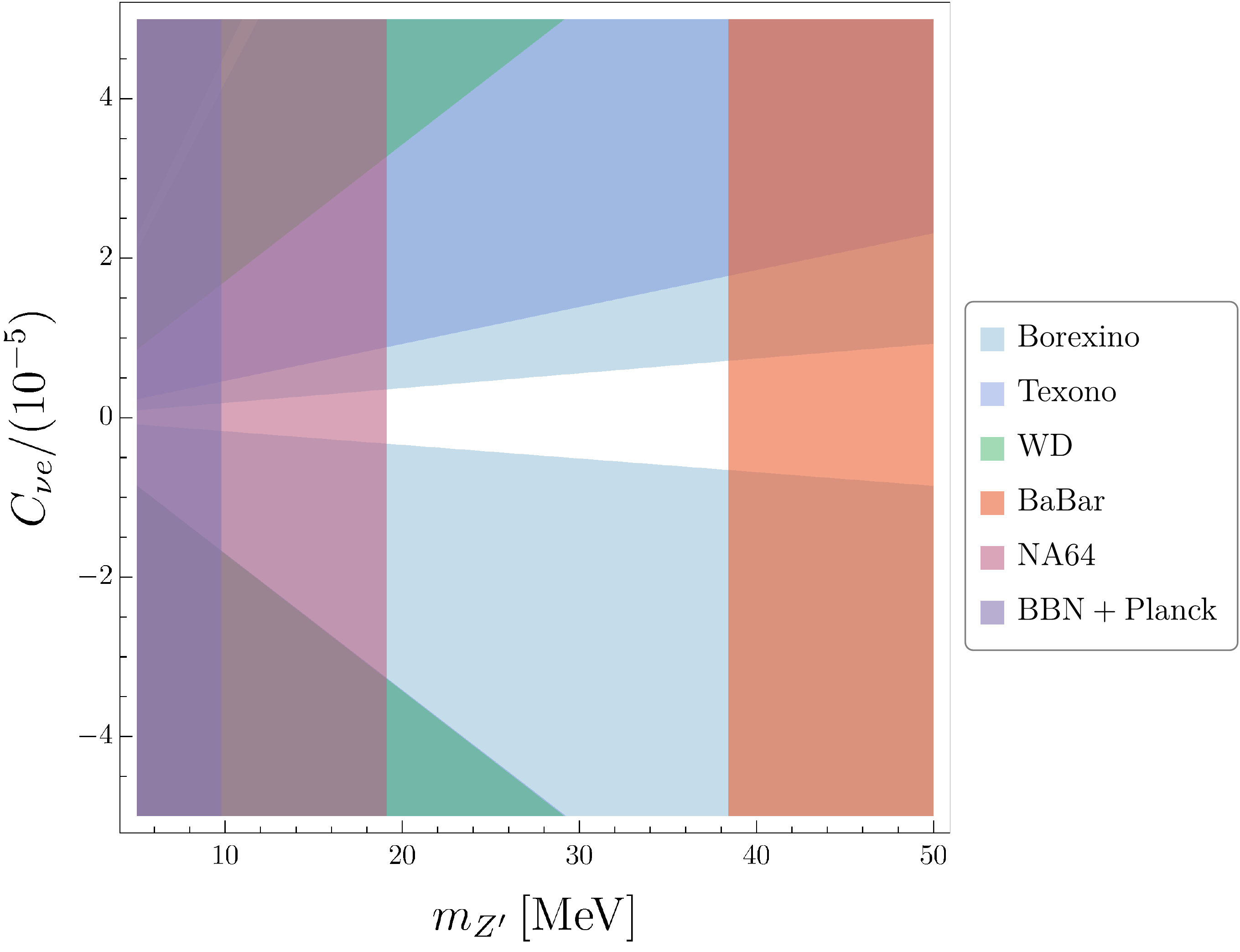}} 
		\caption{\small{Constraints on the mass and effective couplings of the $Z'$ to the electron sector. In the upper plots, we have set $C_{\nu e} = 0$ and taken 
				$|C_{Ve}|$ so that the contribution of the $Z'$ loop induces a value of $a_e$ which is (a) $1\sigma$ below, and (b) $1 \sigma$ above the experimental value. The shaded regions are excluded and in each plot a thin white strip of allowed parameter space remains, indicated by the red arrow. In the lower plots the allowed neutrino coupling is shown, with zero vector coupling $C_{Ve}=0$, and axial coupling $C_{Ae}>0$ such that the contribution of the $Z'$ loop induces a value of $a_e$ which is (c) $1\sigma$ below, and (d) $1 \sigma$ above the experimental value.}}
		\label{fig:e_constraints}
	\end{figure}

	\subsubsection{Analysis of constraints in the electron sector}
	
	We now combine all the constraints discussed above to analyse viable parameter space for the explanation of $\Delta a_{e}$. Our results are summarised in Fig.~\ref{fig:e_constraints}.
	In the plots, effective couplings to muons and taus are set to zero, which is relevant for the bounds from White Dwarfs and Borexino, cf. Eqs.~\eqref{wdconstraint} and \eqref{borexino} respectively. 
	We first set the neutrino coupling, $C_{\nu e}$, to zero in Figs.~\ref{fig:e_constraints}~(a,b) to analyse limits solely on the electron couplings, and plot constraints on the axial electron coupling, $C_{Ae}$, against the vector boson mass, $m_{Z'}$.
	We focus on the axial coupling for two reasons. Firstly, $C_{Ae}$ generates the $\Delta a_e < 0$ required by experiment. 
	Secondly, axial couplings provide various model-building challenges, see Section \ref{sec:models}.

	For each value of $C_{Ae}$ and $m_{Z'}$, in Figs.~\ref{fig:e_constraints}~(a,b) we choose $|C_{Ve}|$ such that the $Z'$ loop induces a correction to $a_e$ that is respectively $1\sigma$ less than and $1\sigma$ greater than the discrepancy of Eq.~\eqref{electronanomaly}, i.e. $\Delta a_e = -12.3 \times 10^{-13}$ in Fig.~\ref{fig:e_constraints}~(a) and $\Delta a_e = -5.1 \times 10^{-13}$ in Fig.~\ref{fig:e_constraints}~(b). 
	This therefore displays the full range of $Z'$ masses and axial couplings which can reduce the $a_e$ anomaly to less than $1\sigma$. 
	Note that the signs of $C_{Ae}$ and $C_{Ve}$ are irrelevant for Figs.~\ref{fig:e_constraints}~(a,b) since all constraints in these plots bound only their absolute values. 
	The blue triangular regions in the lower right half of Figs.~\ref{fig:e_constraints}~(a,b) correspond to values of $C_{Ae}$ and $m_{Z'}$ such that it is impossible to generate the desired deviation in $a_e$, regardless of the value of $C_{Ve}$. 
	The other shaded regions are excluded by the experimental constraints discussed above. 
	In both plots there is a thin white strip, bounded between the blue $\Delta a_e$ and yellow M\o ller exclusion regions, which represents the allowed parameter space. 
	The smallness of these allowed regions shows that even before any model-building considerations are taken into account, it is rather difficult to satisfy the $(g-2)_e$ anomaly while obeying the copious experimental constraints we have mentioned. 
	In the white strips, $|C_{Ve}| < |C_{Ae}|$. 
	Indeed, since the parity-violating M\o ller scattering bound is $|C_{Ve} C_{Ae}| \lesssim 10^{-8}$, we find that since $|C_{Ae}| \gtrsim 1.3 \times 10^{-4}$ is needed to explain the anomalies, we therefore have $C_{Ve} \lesssim 7.7 \times 10^{-5}$.
	As the vector coupling of $Z'$ to electron is required to be smaller than the axial coupling, Eq.~\ref{gm2s} for electron is well approximated by
	\begin{align}\label{reduced_gm2_e}
		\Delta a_e \simeq - 5 \, m_e^2 \,C_{Ae}^2\,/ (12\pi^2 \,m_{Z'}^2) \, .
	\end{align}
	Following this conclusion, we set $C_{Ve} = 0$ in Figs.~\ref{fig:e_constraints}~(c,d) to explore the maximum allowed parameter space for the neutrino coupling, $C_{\nu_e}$, against the mass $m_{Z'}$. 
	Similar to before, in the left plot, Fig.~\ref{fig:e_constraints}~(c), we set $C_{Ae}$ such that $a_e$ is $1 \sigma$ below its experimental value, while in the right plot, Fig.~\ref{fig:e_constraints}~(d), we set $a_e$ to $1\sigma$ above it. 
	These values of $C_{Ae}$ can be taken from Eq.~\eqref{reduced_gm2_e}. 
	We take $C_{Ae} >0$ here: if instead $C_{Ae}<0$, Figs.~\ref{fig:e_constraints}~(c,d) look the same but reflected about the $x$-axis, since all bounds are invariant under $C_{\nu e} \to - C_{\nu e}$ and $C_{Ae} \to -C_{Ae}$ when $C_{Ve} = 0$.  
	The Texono and White Dwarf bounds become apparent in these two plots, however we note that both M\o ller scattering and KLOE constraints are satisfied when $C_{Ve} = 0$ and therefore do not appear. 
	A key conclusion from Figs.~\ref{fig:e_constraints}~(c,d) is that NA64 and BaBar effectively restrict the mass range of a $Z'$ which can satisfy the $a_e$ anomaly to within $1\sigma$ to be $16 \, \mathrm{MeV} \, \lesssim m_{Z'} \lesssim 38 \, \mathrm{MeV}$. 
	In both cases, Borexino gives the strongest constraint on neutrino coupling, $|C_{\nu e}| \lesssim 10^{-5}$, more than an order of magnitude smaller than the required axial coupling.

	In summary, it is clear that a $Z'$ solution to just the $(g-2)_e$ anomaly alone requires some specific ingredients. 
	In particular, $|C_{Ae}| \sim O(10^{-4})$ should be larger than $|C_{Ve}|$ and at least an order of magnitude larger than $|C_{\nu e}|$, 
	while the mass of $Z'$ is constrained to a small window $16$ MeV $\lesssim m_{Z'} \lesssim$ $38$ MeV.

	\subsection{Couplings to muons}
	\label{muoncouplings}
	Now we turn to the bounds on the effective couplings of the $Z'$ to muons and the muon neutrino, namely $C_{V\mu}$, $C_{A\mu}$, and $C_{\nu \mu}$. 
	There are fewer bounds on these than on the couplings to electrons for a few reasons. 
	One is that electrons, being stable, are far easier to handle experimentally. 
	Another reason is that we are led to probe $Z'$ masses sufficiently light that they don't decay into muons. 
	Then, as we have seen, various experiments constrain $C_{Ve,Ae}$ from the absence of $Z' \to e^+ e^-$ but cannot similarly constrain $C_{V\mu,A \mu}$ from the absence of the $Z' \to \mu^+ \mu^-$ decays as these are already kinematically forbidden. 
	Despite this, there remain various strict limits on $Z'$ interactions with muons and muon neutrinos.

	\subsubsection{Cosmological and astrophysical bounds}

	When $|C_{\nu \mu}| \gtrsim 10^{-9}$, bounds from BBN and Planck studied by \cite{Sabti:2019mhn} set a lower limit on the $Z'$ mass, $m_{Z'} \gtrsim 8.3$ MeV. 
	This is similar to the limit on new electrophilic species outlined at the start of Section~\ref{electroncouplings}. 
	For $|C_{\nu \mu}| \lesssim 10^{-9}$, however, it may seem that lower $m_{Z'}$ masses are in principle allowed. 
	For a light $Z'$ ($m_{Z'} \ll m_\mu$), a minimum vector coupling to muons of $|C_{V\mu}| \gtrsim 4\times 10^{-4}$ is required to reduce the $(g-2)_\mu$ tension to within $1\sigma$. 
	This coupling ensures that the $Z'$ was in thermal equilibrium with the SM at earlier times. After decoupling (at temperature $T \sim m_\mu/10$), a very light $Z'$ would constitute an extra relativistic species contributing to the expansion rate of the Universe during neutrino decoupling and BBN, which took place between $10 \,{\rm keV} \lesssim T \lesssim 2 \, {\rm MeV}$. In general, we therefore consider $m_{Z'} \gtrsim 2$ MeV to avoid constraints from measurements of $N_{\rm eff}$ and primordial element abundances.

	Additionally, we note that a study of energy loss in supernovae due to $Z'-\mu$ interactions by \cite{Croon:2020lrf} rules out a $Z'$ with coupling $|C_{V\mu}| \gtrsim 4\times 10^{-4}$ for masses less than $\mathcal{O}(100)$ eV.\footnote{In the models studied in \cite{Croon:2020lrf}, the $Z'$ has interactions with both muons and muon neutrinos. However, at low masses $(\ll$ MeV) it is the $Z'-\mu$ interactions with dominate the bounds, while the $Z'-\nu_\mu$ interaction plays a negligible role. } 
	Recall, however, that for $m_{Z'}\gtrsim 100$ eV, the effective coupling required to explain the $(g-2)_{e}$ anomaly must be greater than $10^{-9}$. 
	With an interaction of this size, the BBN bound on a new electrophilic species dictates that $m_{Z'}$ must be at least in the MeV range.

	We can therefore rule out the possibility of an extremely light $Z'$ (i.e. $m_{Z'} \ll$ MeV) being able to explain the two $g-2$ anomalies. 
	Its mass must consequently be at least 16 MeV, as we showed from the analysis of constraints on $Z'$ couplings to the electron sector in the previous section.

	\subsubsection{Neutrino scattering bounds}

\begin{figure}[hbt!]
		\subcapraggedrighttrue
		\subfigure[\footnotesize{$C_{\nu \mu} = 10^{-5}$, $C_{A\mu}$ fixed such that the $a_\mu$ anomaly is exactly satisfied.}]{\includegraphics[height=0.36\textwidth, width=0.48\textwidth]{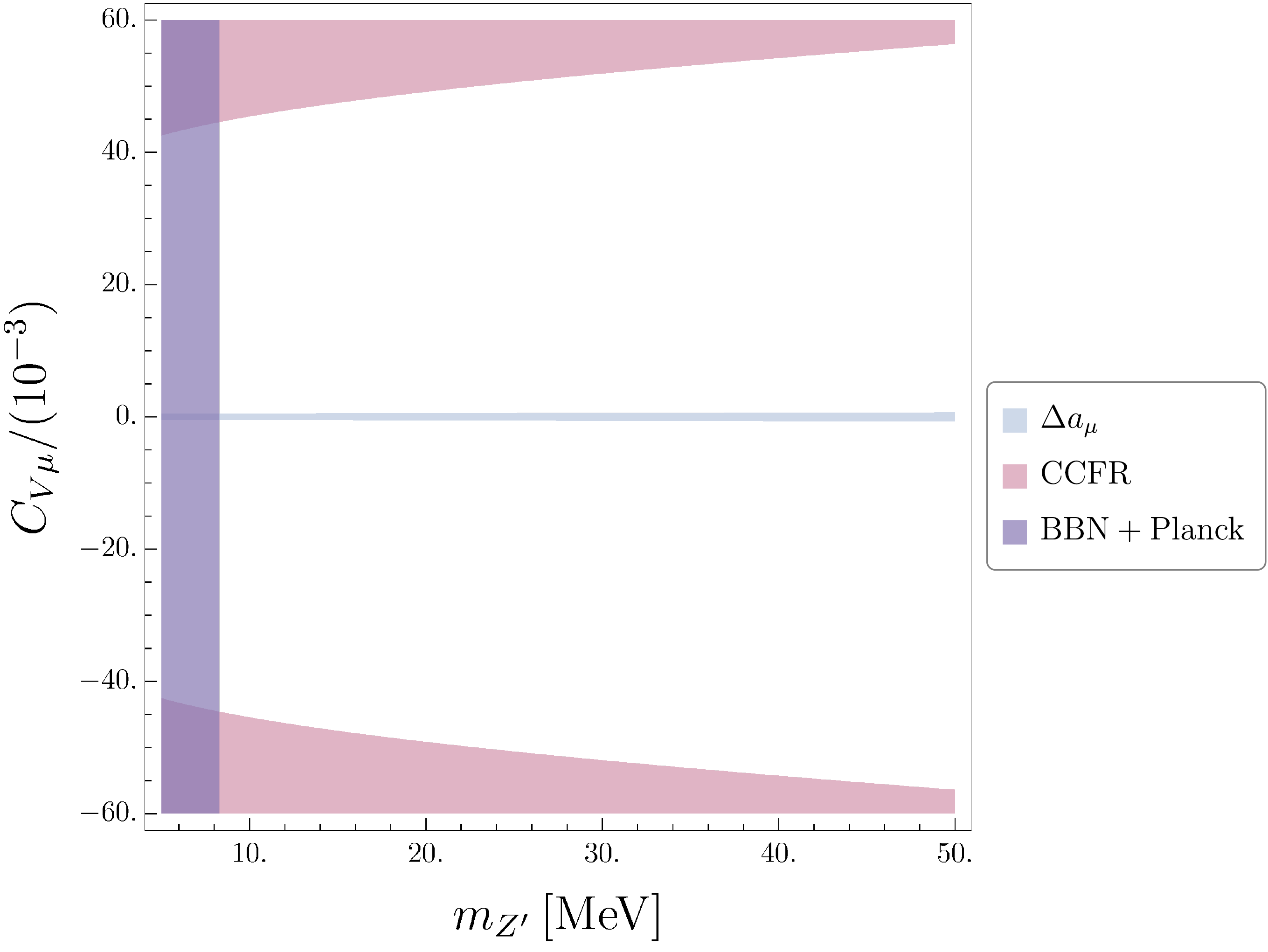}} \quad
		\subfigure[\footnotesize{Contours of $|C_{A \mu}|$ such that the $a_{\mu}$ anomaly is exactly satisfied.}]{\includegraphics[height=0.36\textwidth,width=0.48\textwidth]{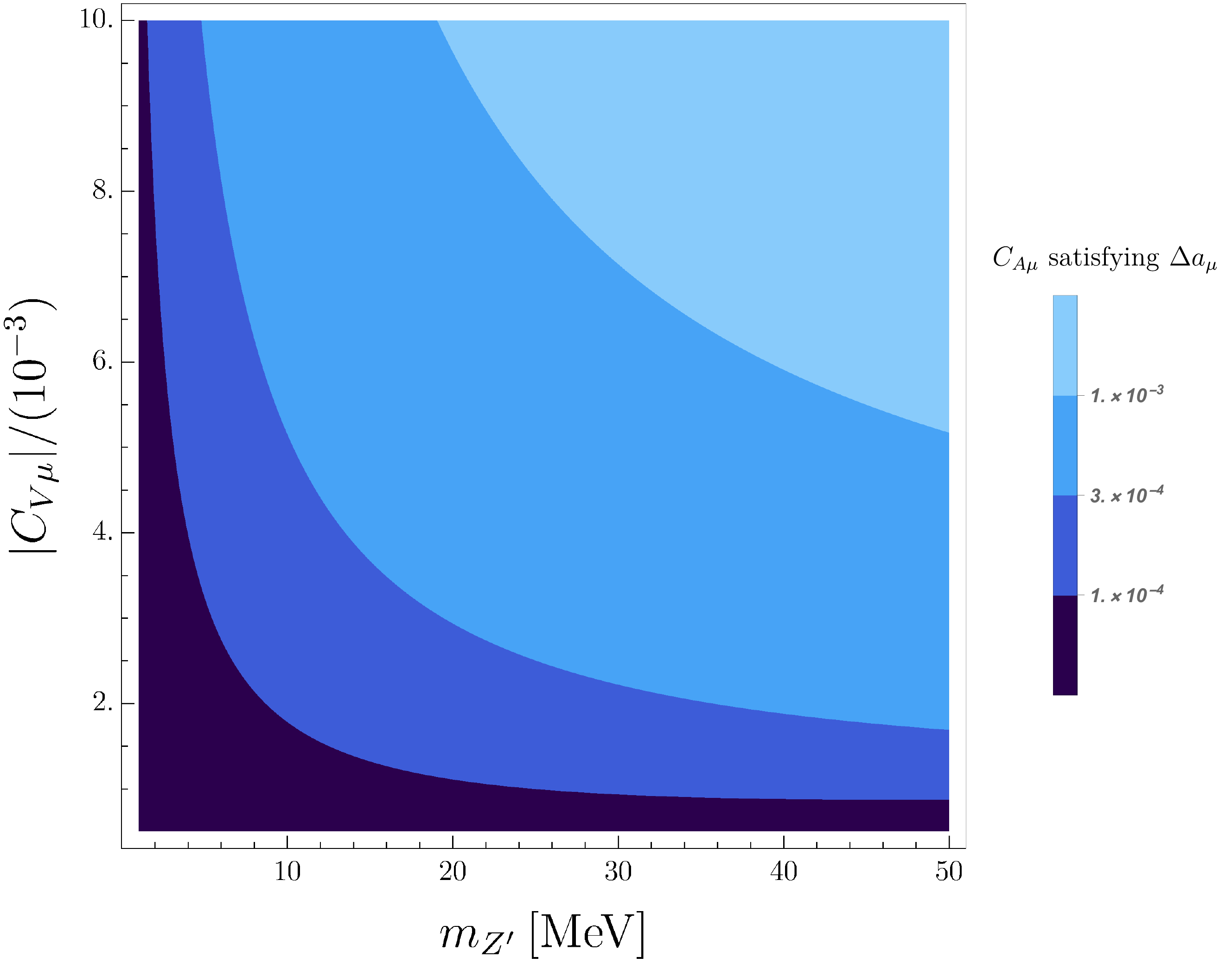}} \quad
		\subfigure[\footnotesize{$C_{V \mu}, C_{A \mu} = 0$, $C_{A e}>0$ fixed such that $a_{e}$ is $1 \sigma$ below the experimental value.}]{\includegraphics[height=0.36\textwidth,width=0.48\textwidth]{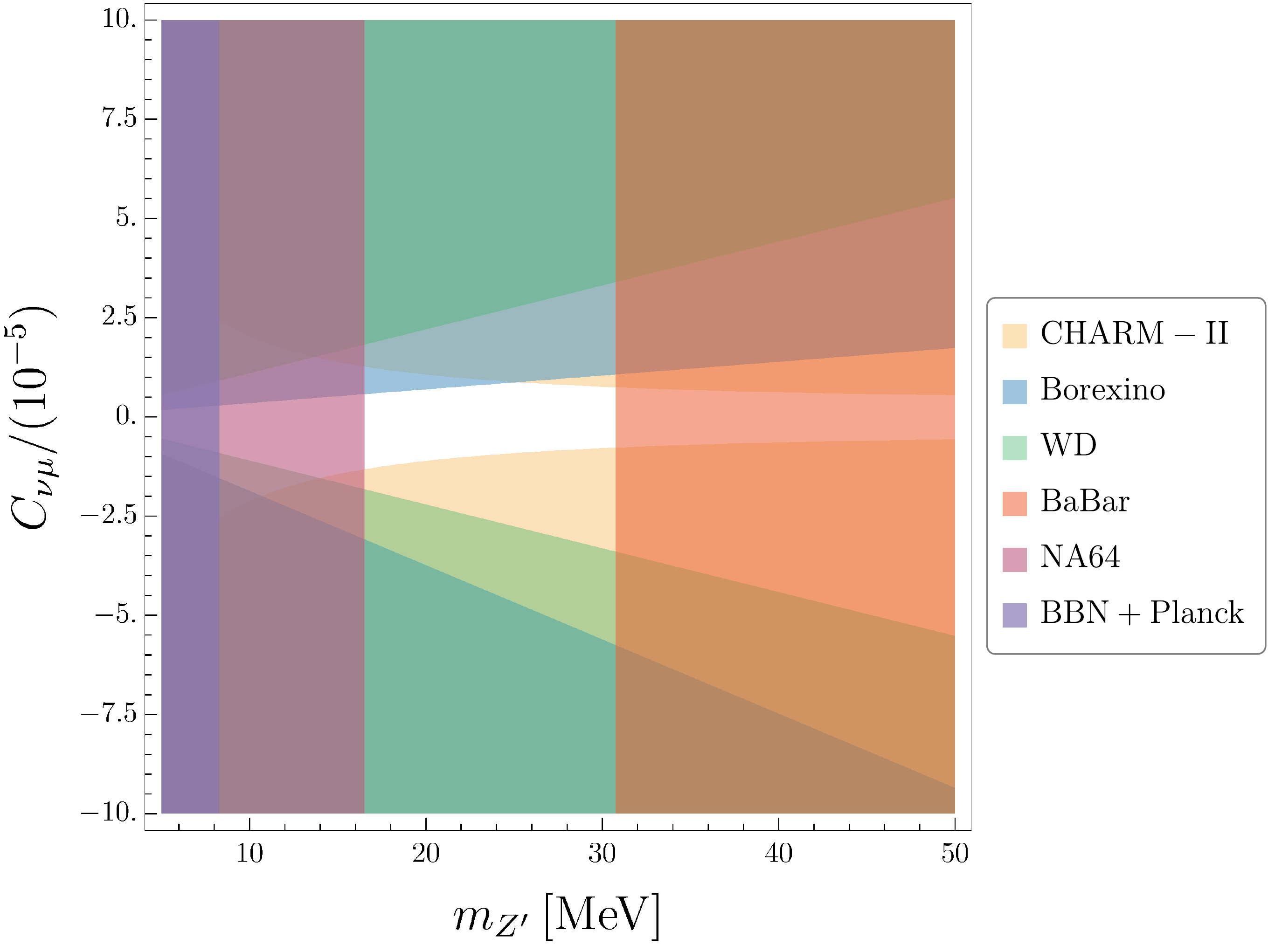}} \quad
		\subfigure[\footnotesize{$C_{V \mu}, C_{A \mu} = 0$, $C_{A e}>0$ fixed such that $a_{e}$ is $1 \sigma$ above the experimental value.}]{\includegraphics[height=0.36\textwidth,width=0.48\textwidth]{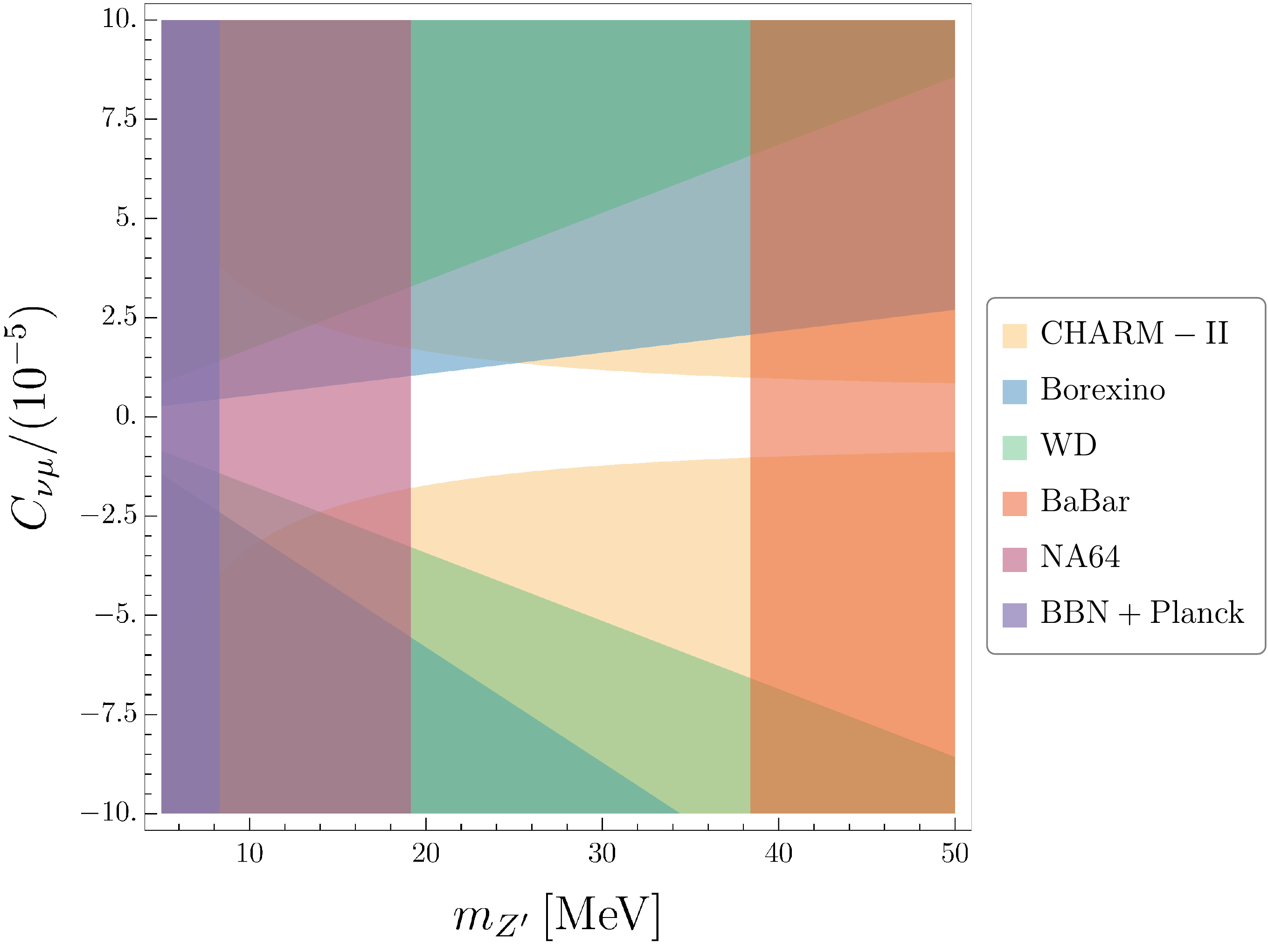}} \quad
		\caption{\small Constraints on the mass and effective couplings of the $Z'$ in the muon sector. In (a) we have set $C_{\nu \mu} = 10^{-5}$ fixed $C_{A\mu}$ so that the $a_\mu$ anomaly is exactly satisfied, while in (b) we show contours of the values of $|C_{A\mu}|$ this corresponds to, as a function of $|C_{V\mu}|$ and $m_{Z'}$. 
			In the bottom two plots we focus on the neutrino couplings, setting $C_{Ae} >0$ and $C_{Ve} = 0$ such that the contribution of the $Z'$ to $a_e$ is (c) $1\sigma$ below, and (d) $1\sigma$ above the experimental value. See text for more details.
		}
		\label{fig:mu_constraints}
	\end{figure}

	Several neutrino scattering experiments bound couplings to muons and muon neutrinos. 
	The most stringent of these are Borexino and CHARM-II, introduced above. 
	The Borexino result was given in Eq.~\eqref{borexino}. 
	The mean (anti)neutrino energy in the CHARM-II experiment is much larger than the $Z'$ masses we consider, with $\langle E_\nu \rangle = 23.7$ GeV and $\langle E_{\bar{\nu}} \rangle = 19.1$ GeV \cite{Vilain:1993kd}, therefore the approximation $m_{Z'} \gg \sqrt{m_e T}$ which we used to obtain Eqs.~\eqref{texono} and \eqref{borexino} cannot be used. 
	We apply the formalism in \cite{Harnik:2012ni,Lindner:2018kjo} to obtain numerical results, which enter into Fig.~\ref{fig:mu_constraints} by enforcing that the shift in the neutrino scattering cross-section induced by the $Z'$ is no greater than $6\%$ \cite{Lindner:2018kjo}. 
	We mention that some doubts on the CHARM-II analysis were presented in \cite{Bauer:2018onh}, however we do not enter into this discussion.

	A $Z'$ with couplings to muons and muon neutrinos also modifies the neutrino trident process, $\nu_\mu  N \to \nu_\mu \mu^+ \mu^-  N$ \cite{Altmannshofer:2014pba}. 
	Neglecting the coupling $C_{A \mu}$, since $|C_{A \mu}| \ll |C_{V \mu}|$ is necessary to explain the $(g-2)_\mu$ anomaly when $m_{Z'} \lesssim m_\mu$ (see Eq.~\eqref{gm2s}), the trident cross-section including the $Z'$ contribution is \cite{Altmannshofer:2014pba}
	\begin{equation}
		\frac{\sigma_\text{Trident}}{\sigma_\text{Trident}^\text{SM}} \simeq 1 + 5.6 \times 10^5 C_{V\mu} C_{\nu \mu} + 1.3 \times 10^{11} C_{V \mu}^2 C_{\nu \mu}^2 \log \frac{m_\mu^2}{m_{Z'}^2}  ~.
		\label{neutrinotrident}
	\end{equation}
	This can be compared with the CCFR measurement, $\sigma^\text{CCFR}/\sigma^\text{SM} = 0.82 \pm 0.28$ \cite{Mishra:1991bv}, to give a constraint.

	\subsubsection{Analysis of constraints in the muon sector}
	
	%
	
	We combine the results of the above constraints in Fig.~\ref{fig:mu_constraints}. In Fig.~\ref{fig:mu_constraints}~(a) we plot the allowed regions for the effective vector coupling of the $Z'$ to the muon, $C_{V \mu}$, against the $Z'$ mass, with relevant constraints overlaid, while setting $C_{\nu \mu} = 10^{-5}$ and fixing $C_{A\mu}$ such that $(g-2)_{\mu}$ is satisfied. 
	The large white-space shows the parameter space which can explain the $(g-2)_\mu$ anomaly and is not excluded. 
	It reflects the fact that when a light $Z'$ couples only to muons, there are few relevant constraints. 
	$|C_{V\mu}|$ can be as large as $0.05$ in the mass region of interest, $16$ MeV $\lesssim m_{Z'} \lesssim 38$ MeV. 
	In purple on the left of Fig.~\ref{fig:mu_constraints} (a), (c) and (d) is the mass bound $m_{Z'} \gtrsim 8.3$ MeV from BBN and Planck data. 
	The pink bounds at the top and bottom of (a) come from the CCFR neutrino trident measurement, see Eq. \eqref{neutrinotrident}. 
	The thin blue line for $|C_{V\mu}| \lesssim 5\times 10^{-4}$ gives the region for which $|C_{V\mu}|$ is too small to reduce the $(g-2)_\mu$ tension to within $1\sigma$, even if $C_{A\mu} = 0$. 
	Then (b) shows contours of the necessary values of $|C_{A\mu}|$ to exactly satisfy the anomaly, given $m_{Z'}$ and $C_{V\mu}$. This can be as large as $\mathcal{O}(10^{-3})$, but must always be at least factor of a few smaller than $|C_{V\mu}|$.

	The bounds on the $Z'$ interaction with muon neutrinos are significantly stronger. 
	Fig.~\ref{fig:mu_constraints} (c) and (d) show bounds on neutrino couplings from various experiments (we take $C_{\nu e} = C_{\nu \tau} = 0$). 
	We must invoke couplings to electrons, since modifications to both neutrino scattering on electrons and white dwarf cooling necessarily depend on the $Z'$ coupling to electrons, as does $Z'$ detection at beam dumps.  
	To be as minimal as possible, we take only non-zero $C_{Ae}$, assuming $C_{Ve} = 0$. 
	In (c) $C_{Ae} > 0$ is set (as a function of $m_{Z'}$) such that $a_e$ is $1\sigma$ below its experimental value, while in (d) it is set such that $a_e$ is instead $1\sigma$ above. 
	This allows us to see the full range of allowed $C_{\nu \mu}$. 
	Clearly, its absolute value cannot be much larger than $\sim 2 \times 10^{-5}$, which justifies the choice of $C_{\nu \mu}$ in plot (a). 
	Taking $C_{Ae} < 0$ instead would only flip (c) and (d) about the $x$-axis, since the neutrino scattering and white dwarf constraints are invariant under $C_{Ae} \to -C_{Ae}$ and $C_{\nu \mu} \to -C_{\nu \mu}$ when those are the only non-zero couplings.

	\subsection{Future Discovery Potential}
	\label{sec:future}
	Having surveyed the current limits, in this section we will discuss future experiments which could discover (or preclude) the low scale $Z'$ explanation of $\anomalies$ by closing the allowed parameter space given in Figs.~\ref{fig:e_constraints} and \ref{fig:mu_constraints} (keeping in mind that these were generated assuming the Caesium $a_e$ result). The place to start is with the magnetic dipole moment anomalies themselves. 
	The two highly inconsistent measurements of $\alpha_\textrm{em}$ (from which the value of $(g-2)_e$ is derived) made in Caesium \cite{Parker:2018vye} and Rubidium \cite{Morel:2020dww} atoms demand a third independent experiment to resolve the situation. 
	It is indeed not even clear whether an anomaly exists. 
	On top of this, the Muon g-2 and J-PARC experiments \cite{Grange:2015fou,Abe:2019thb} are expected to provide improved measurements of $a_\mu$, which is particularly important given the recent debate about the SM prediction \cite{Borsanyi:2020mff}. 
	Beyond this, there are several future experiments which are expected to test the allowed $Z'$ couplings to charged leptons.

	We note first of all that an improved measurement of parity-violating M\o ller scattering can never close the parameter space, as this bounds the combination $|C_{Ve} C_{Ae}|$, which can always be satisfied by taking one of $C_{Ve}$ or $C_{Ae}$ to zero while the other (depending on the sign of the $a_e$ anomaly) explains the discrepancy. Thus, we will not discuss future experiments in this area.

	To fully probe the available space, we require other bounds to be strengthened. Currently, the lower bound on the $Z'$ mass, $m_{Z'} \gtrsim 16$ MeV, is fixed from NA64's visible decay limits. 
	Future NA64 results after the LHC's Long Shutdown 2 (LS2) should probe masses up to around 20 MeV \cite{2003.07257}.
	
	For higher $Z'$ masses, the most sensitive future experiment will be Belle-II \cite{1808.10567}, from the visible dark photon search mode $A' \rightarrow \ell \ell$. With 50 ab${}^{-1}$ luminosity (expected in 2025), the projected sensitivity is \cite{Ferber:2015jzj}
	\begin{equation}
		\sqrt{(C_{Ve}^2 + C_{Ae} ^2) \textrm{BR}(Z' \rightarrow e^+ e^-)} \lesssim 9 \times 10^{-5} .
	\end{equation}
	An alternative experiment with similar sensitivity is MAGIX at MESA \cite{1809.07168}, which is also currently under construction and expecting results in the next few years.
	The combination of NA64 and Belle-II (or MAGIX) could entirely rule out or discover low scale $Z'$ explanations of the current Caesium $(g-2)_e$ result. 
	Beam dumps (e.g. FASER \cite{1811.12522} and SHiP \cite{1504.04855}) are also expected to play a role. 
	This provides hope that a firm conclusion could be reached within the next few years.

	The MUonE experiment \cite{Banerjee:2020tdt} will probe the product of couplings to electrons and to muons. 
	In this way it is a unique test of a $Z'$ which explains both anomalies, because it is required to have significant couplings to both leptons. 
	The experiment is expected to cover a significant portion of the parameter space which remains open, see \cite{Dev:2020drf,Masiero:2020vxk}.

	Finally, we point out that while there are many dark photon experiments beyond those listed above, many do not directly test our framework. There are two reasons for this. Firstly, we are concerned with the lepton-$Z'$ couplings only, so experiments which involve production of the $Z'$ though quarks are not applicable. This includes electron-proton scattering (such as DarkLight \cite{1412.4717}), proton-proton scattering and pion decays (e.g. NA62 \cite{CortinaGil:2019nuo}). 
	Secondly, we require visible ($Z' \rightarrow ee$) decays of the $Z'$, which excludes the invisible-only experiments such as  PADME \cite{KozhuharovonbehalfofthePADMECollaboration:2018exf}, VEPP-3 \cite{Wojtsekhowski:2017ijn}, BDX \cite{Battaglieri:2019nok} and LDMX \cite{1808.05219}. Consequently, the available parameter space in Figs.~\ref{fig:e_constraints} and \ref{fig:mu_constraints}, and hence the discovery potential, may only be fully reached by the small number of experiments which focus on vector bosons produced by leptons and which decay to $e^+ e^-$.

	\section{Viability of specific $Z'$ models}
	\label{sec:models}
	Having completed our model-independent analysis in Section \ref{sec:constraints}, we now turn to specific realisations of $Z'$ models. 
	The ingredients for the simultaneous explanation of the $(g-2)$ anomalies with a single $Z'$ are:
	\begin{enumerate}
		\item A light $Z'$ in the mass range $16 \; {\rm MeV} \lesssim m_{Z'} \lesssim 38 \; {\rm MeV}$.
		
		\item Axial coupling of the $Z'$ to electrons larger than the vector coupling: $|C_{Ae}| \sim [1-3.2] \times 10^{-4}, \, |C_{Ve}| \lesssim 7.7\times 10^{-5}$.
		
		\item Large vector coupling to muons, $5 \times 10^{-4} < |C_{V\mu}| \lesssim 0.05$, and an axial coupling $C_{A\mu}$ that is smaller by at least a factor of a few.
		
		\item Tiny $Z'$ couplings to neutrinos: $|C_{\nu_e,\, \nu_\mu}| \lesssim 10^{-5}$.
	\end{enumerate}
	
	We now attempt to realise this hierarchy of couplings in various classes of $Z'$ models, each of which inevitably introduces additional relations between effective couplings. 
	We will begin with the simplest case of just the SM extended by a $U(1)'$. We will then move onto a scenario with additional Higgs doublets, and finally discuss the viability of a Froggatt-Nielsen style model, in which the gauge invariance of the charged lepton Yukawa interactions is relaxed.
	Note that in each case the dominant contribution to the shift in $(g-2)_{e,\mu}$ comes solely from the $Z'$.

	Before commencing, we also remark that the cancellation of gauge anomalies is crucial for constructing a consistent theory. The $U(1)'^3$ and $U(1)'\text{grav}^2$ anomalies can always be satisfied by introducing additional chiral fermions which are charged under the $U(1)'$ but sterile with respect to the SM (in fact one needs at most five \cite{Allanach:2019uuu}). The anomaly cancellation conditions involving SM groups are typically more challenging to satisfy. However, this section addresses the primary question of whether it is possible to generate the desired effective couplings, without delving into how to do so in an anomaly-free way.

	\subsection{SM+$U(1)'$}
	\label{u1model}

	First consider a minimal $Z'$ model, in which the SM is extended by a gauge $U(1)'$ and also add a scalar, $S$, charged under the $U(1)'$, whose non-zero VEV, $\langle S \rangle = v_S/\sqrt{2}$, spontaneously breaks the $U(1)'$ symmetry. 
	We note here that this unspecified $U(1)'$ covers in particular the case of gauging combinations of electron, muon and tau number, i.e. $U(1)_{x e + y\mu + z\tau}$ for some $x,y,z$. 
	Let us establish the formalism, which will also be useful for the subsequent models.

	In general there is mixing between $U(1)_Y$ and $U(1)'$, and the kinetic terms for the pair of $U(1)$s can be written as
	\begin{equation}
		\mathcal{L}_\text{kin} \supset - \frac{1}{4} B'_{\mu \nu} B'^{\mu \nu} - \frac{1}{4} X'_{\mu \nu} X'^{\mu \nu} - \frac{\varepsilon}{2} B'_{\mu \nu} X'^{\mu \nu} ~,
		\label{kinmix}
	\end{equation}
	where $X'_\mu$ is the gauge field associated with $U(1)'$ and $X'_{\mu \nu}$ is the corresponding field strength tensor.

	An appropriate rotation and rescaling of fields removes the mixing (see e.g. \cite{Coriano:2015sea}),
	\begin{equation}
		\begin{pmatrix}
			B'_{\mu}\\
			X'_{\mu}
		\end{pmatrix} = 
		\begin{pmatrix}
			1 & \frac{-\varepsilon}{\sqrt{1-\varepsilon^2}}\\
			0 & \frac{1}{\sqrt{1-\varepsilon^2}}
		\end{pmatrix}
		\begin{pmatrix}
			B_{\mu}\\
			X_{\mu}
		\end{pmatrix} \, ,
	\end{equation}
	and leaves the couplings in the covariant derivative in the form, 
	\begin{equation}
		D_\mu = \partial_\mu + i g_1 Y B_\mu + i (\tilde{g} Y + g' z) X_\mu ~,
		\label{modifiedcovdev}
	\end{equation}
	where $g'$ and $g_1$ are the respective $U(1)'$ and $U(1)_Y$ gauge couplings, and $Y$ and $z$ are the respective charges of the field under $U(1)_Y$ and $U(1)'$. 
	In the above, we have only kept terms that are leading order in the kinetic mixing parameter $\varepsilon$, which is taken to be small. This gives $\tilde{g} \simeq -g_1 \varepsilon$. 
	Breaking the EW and $U(1)'$ symmetries and diagonalising the gauge boson mass matrix, we move into the basis of mass eigenstates, $A_\mu$, $Z_\mu$, and $Z'_\mu$, using
	\begin{equation}
		\begin{pmatrix}
			B_{\mu}\\
			W_{\mu}^{3}\\
			X_{\mu}
		\end{pmatrix} =
		\begin{pmatrix}
			c_w & -s_w c_{\phi} & s_w s_{\phi}\\
			s_w & c_w c_{\phi} & -c_w s_{\phi}\\
			0 & s_{\phi} & c_{\phi}
		\end{pmatrix}
		\begin{pmatrix}
			A_{\mu}\\
			Z_{\mu}\\
			Z'_{\mu}
		\end{pmatrix} \, ,
	\end{equation}
	where $w$ is the weak-mixing angle, $\phi$ is the $Z-Z'$ mixing angle, and $s$ ($c$) denotes sine (cosine). 
	This gauge boson mixing is given by
	\begin{equation}
		\tan 2\phi \simeq \frac{2z'_H g' e/(s_w c_w)}{z'^2_H g'^2 + (2 z_S g' v_S/v)^2 - e^2/(s_w^2 c_w^2)} ~,
		\label{zzpmixing}
	\end{equation}
	where $z'_H \simeq \tilde{g}/g' + 2 z_H$, $z_H$ ($z_S$) is the $U(1)'$ charge of the Higgs $(S)$.
	Finally, after outlining this procedure, we can write the effective couplings of SM fermions to the gauge boson mass eigenstates. 
	We find that the effective couplings for charged leptons at leading order in $g', \tilde{g}$ are
	\begin{align}
		C_{V\alpha} &\simeq -\tilde{g} c_w^2 + \frac{g'}{2} ( z_H (4 s_w^2 - 1) + z_{L\alpha} + z_{R\alpha}) ~, \label{axialCouplingSMU1a} \\
		C_{A\alpha} &\simeq \frac{g'}{2} ( z_H + z_{R\alpha} - z_{L\alpha}) ~,
		\label{axialCouplingSMU1b}
	\end{align}
	where $z_{L\alpha}$ ($z_{R\alpha}$) is the $U(1)'$ charge of the lepton doublet (singlet), $l_{L\alpha}$ ($e_{R\alpha}$).

	Here we see from the $U(1)'$ invariance of the SM charged lepton Yukawa couplings, $\mathcal{L} \supset - \overline{l_L} Y_e H e_R + h.c.$, we have that $z_{L\alpha} = z_{R \alpha} + z_H$. 
	Consequently, $C_{A\alpha} = 0$ at leading order in $g', \tilde{g}$, and therefore $|C_{V\alpha}| \gg |C_{A\alpha}|$. 
	Under this condition, the $Z'$ with $m_{Z'} > m_e$ always induces a positive shift in $a_e$, cf. Eq.~\eqref{gm2s}, which is the wrong direction for explaining the Caesium anomaly. 
	The simplest $U(1)'$ extension of the SM can therefore be ruled out as a possibility of resolving both $(g-2)_e$ and $(g-2)_\mu$ discrepancies.

	\subsection{NHDM$+U(1)'$}

	We have seen that extending the SM by just a gauge $U(1)'$ and a scalar does not give us enough freedom to arrange $|C_{Ae}| \gtrsim |C_{Ve}|$. 
	There are several options to circumvent this problem, including a) introducing new fermions which mix with the SM ones, b) extending the Higgs sector, or c) removing the gauge invariance of the Yukawas via a Froggatt-Nielsen \cite{Froggatt:1978nt} type set-up. 
	In the case of option (a), our analysis is not valid because loops involving the new fermions could also contribute to $\anomalies$.\footnote{An attempt to explain both anomalies by introducing a heavy vector-like fourth family of leptons was made in \cite{CarcamoHernandez:2019ydc} but was ultimately unsuccessful.} 
	Here we consider option (b).
	This was previously explored e.g. in the context of the Atomki anomaly \cite{DelleRose:2017xil}. 
	In Section \ref{fntype} we will consider option (c).

	Let us take the type-I 2HDM, wherein all SM fermions couple to the same Higgs doublet, $H_2$. 
	This choice will not be important for the following discussion, since we are concerned only with the lepton couplings, thus our discussion is general. 
	We can also generalise to the case of many Higgs doublets, see for instance Appendix A of \cite{Lindner:2018kjo}. 
	The key point is that this set-up modifies Eq.~\eqref{axialCouplingSMU1b} and therefore permits non-negligible axial couplings.

	The kinetic mixing between $U(1)_Y$ and $U(1)'$ and the subsequent modification of covariant derivatives is as described in Eqs.~\eqref{kinmix}-\eqref{modifiedcovdev}. 
	The neutral gauge boson mass mixing is modified by the presence of two Higgs fields, $H_{1,2}$, with $U(1)'$ charges $z_{1,2}$ and VEVs $\langle H_{1,2} \rangle = (0~,~ v_{1,2}/\sqrt{2})^T$, where $v_1 = v \cos \beta$ and $v_2 = v \sin \beta$. 
	Then the mixing angle is given by 
	\begin{equation}
		\tan 2 \phi \simeq \frac{2z_H g' e/(s_w c_w)}{z_{H^2}^2 g'^2 + (2 z_S g' v_S/v)^2 - e^2/(s_w^2 c_w^2)} ~,
		\label{zzpmixing2hdm}
	\end{equation}
	where
	\begin{align}
		& z_H = z_1' c_\beta^2 + z_2' s_\beta^2 ~,&
		& z_{H^2} = z_1'^2 c_\beta^2 + z_2'^2 s_\beta^2 ~,&
	\end{align}
	with $z_j' = \tilde{g}/g' + 2 z_j$ for $j = 1,2$. 
	Note that in the limit $\beta \to 0 ~(\pi)$, i.e. when only $v_1$ ($v_2$) is non-zero, we recover the result of Eq.~\eqref{zzpmixing} up to $z_H \to z_1 ~(z_2)$. 
	Accounting for the kinetic and mass mixing, the effective couplings for charged leptons and neutrinos at leading order in in $g', \tilde{g}$ are
	\begin{align}
		C_{V\alpha} &\simeq z_{L \alpha} g' - c_w^2 \tilde{g}- \frac{g'}{2} \left[ (1 - 4s_w^2) c_\beta^2 z_1 + ( 1 + s_\beta^2 - 4s_w^2 s_\beta^2 ) z_2 \right] \\
		C_{A \alpha} &\simeq \frac{(z_1 - z_2)}{2} c_\beta^2 g'  \\
		C_{\nu \alpha} &\simeq - \frac{\tilde{g}}{2} + g' (z_{L \alpha} + \frac{z_H}{2}) ~,
	\end{align}
	using that the $U(1)'$-invariance of the charged lepton Yukawa couplings demands $z_{R\alpha} = z_{L\alpha} - z_2$. 
	We see that $C_{A \alpha}$ can be non-zero when $z_1 \neq z_2$, and that it is flavour-universal. 
	$C_{V\alpha}$ and $C_{\nu \alpha}$, on the other hand, are flavour-dependent. 
	However, both depend linearly on $z_{L \alpha}$, so that
	\begin{equation}
		C_{Ve} - C_{V \mu} = g' (z_{Le} - z_{L\mu}) = C_{\nu e} - C_{\nu \mu} ~.
		\label{cVdifference}
	\end{equation}
	Consequently, there are not six independent effective couplings $C_{V\alpha}, C_{A \alpha}, C_{\nu \alpha}$ for $\alpha = e,\mu$, but rather only four are independent.
	Given this, it is in fact simple to argue that this class of models cannot simultaneously explain the $\anomalies$ anomalies. 
	Our model-independent analysis in Section \ref{sec:constraints} established that due to the stringency of the bounds from neutrino scattering experiments, the effective neutrino couplings must be tiny: $C_{\nu e}, C_{\nu \mu} \lesssim 10^{-5}$, cf. Figs.~\ref{fig:e_constraints} and \ref{fig:mu_constraints}. 
	From Eq.~\eqref{cVdifference}, this implies that we need $|C_{Ve} - C_{V\mu}| \lesssim 10^{-5}$. 
	However, it is apparent from points 2 and 3 of the summary list at the beginning of this section that $|C_{Ve} - C_{V\mu}| \gtrsim 4 \times 10^{-4}$. Clearly, this framework is not successful.

	In the simplest $U(1)'$ extension of the SM, only the $(g-2)_\mu$ anomaly could be resolved as it was impossible to generate significant axial couplings of the $Z'$. 
	Introducing additional Higgs fields enables large axial couplings, so that either the $(g-2)_e$ or the $(g-2)_\mu$ anomaly may be explained. 
	However, the correlations between different effective couplings and the strength of the bounds on neutrino couplings conspire to preclude an explanation of both anomalies at the same time.

	\subsection{Froggatt-Nielsen model}
	\label{fntype}
	A second way to generate sizeable axial couplings, as is necessary to explain the Caesium $(g-2)_e$ anomaly, is by considering a Froggatt-Nielsen type model \cite{Froggatt:1978nt}. 
	In this set-up, we modify the charged lepton Yukawa interactions to some effective interactions of the form,
	\begin{equation}
		\mathcal{L} \supset - \frac{\lambda_{\alpha \beta}}{\Lambda^{n_{\alpha \beta}}} \overline{l_{L\alpha}} H e_{R\beta} \varphi^{n_{\alpha \beta}} + h.c. \, .
		\label{FNlag}
	\end{equation}
	Here $\lambda_{\alpha \beta} = \lambda_\alpha \delta_{\alpha \beta}$ is a diagonal matrix of couplings (in the charged lepton mass basis), $\varphi$ is a flavon, $n_{\alpha \beta} = n_\alpha \delta_{\alpha \beta}$ is a diagonal matrix whose entries are determined by the $U(1)'$ charges of the flavon and the SM leptons, and $\Lambda$ is the scale of some unspecified UV physics. 
	Then the SM charged lepton Yukawa couplings are recovered at the non-zero VEV of the flavon, i.e. $y_\alpha = \lambda_{\alpha} (\langle \varphi \rangle/\Lambda)^{n_{\alpha}}$. 
	More complicated set-ups can also be written down (e.g. the clockwork model of \cite{Smolkovic:2019jow}), and there may be more than one flavon.

	The introduction of flavons removes the relation between the $U(1)'$ charges of the SM leptons and the Higgs. 
	This permits non-vanishing axial $Z'$ couplings at leading order in $g'$, unlike in the standard SM$+U(1)'$ scenario, (recall Eq.~\eqref{axialCouplingSMU1b}). 
	From Eq.~\eqref{FNlag} we have $z_H + z_{R\alpha} - z_{L\alpha} = -n_\alpha$, and we have the freedom to treat $n_\alpha$ as a free, family-dependent parameter.\footnote{In this framework we will not attempt to generate the observed charged lepton Yukawa couplings, but rather focus on whether the $\anomalies$ anomalies can be simultaneously explained} 
	In all other respects, the formalism of this model follows that of the $U(1)'$ extension outlined in Section \ref{u1model}.  Eqs.~\eqref{zzpmixing}-\eqref{axialCouplingSMU1b} still hold, while the effective neutrino couplings are given by
	\begin{equation}
		C_{\nu \alpha} \simeq g' (z_H + z_{L\alpha}) \, ,
		\label{neutrinoCouplingsFN}
	\end{equation}
	at leading order in $g',\tilde{g}$. 
	This model was previously studied in \cite{DelleRose:2018eic} to explain the Atomki Beryllium anomaly \cite{Krasznahorkay:2015iga}, another instance in which unsuppressed $C_{Ae}$ is required.

	Combining Eqs.~\eqref{axialCouplingSMU1a}, \eqref{axialCouplingSMU1b} and \eqref{neutrinoCouplingsFN} gives
	\begin{equation}
		C_{Ve} - C_{Ae} - C_{\nu e} = C_{V\mu} - C_{A\mu} - C_{\nu \mu} \, .
		\label{cdifferenceFN}
	\end{equation}
	This is a generalisation of Eq.~\eqref{cVdifference} to the case of non-universal $C_{A}$. 
	However, we see from Fig.~\ref{fig:e_constraints}~(a) and Eq.~\eqref{gm2sfull} that in order to reduce the $\anomalies$ anomalies to $< 1 \sigma$ while satisfying all experimental constraints, we require $|C_{Ae}| \lesssim 3.2 \times 10^{-4}$ and $|C_{V\mu}| \gtrsim 4.8 \times 10^{-4}$, given $m_{Z'} \gtrsim 16$ MeV, the lower bound on the $Z'$ mass obtained in Section \ref{sec:constraints}. 
	However, we have established that $|C_{\nu e}|,|C_{\nu \mu}| \ll 10^{-4}$, while the M\o ller scattering bound gives $|C_{Ve}| \lesssim 3 \times 10^{-5}$ for $|C_{Ae}| \sim 3 \times 10^{-4}$. 
	Finally, a sizeable $|C_{A\mu}|$ demands an even larger $|C_{V\mu}|$, since $\Delta a_\mu \propto (m_{Z'}^2 C_{V\mu}^2 - m_\mu^2 C_{A \mu}^2)$ to a good approximation, see Eqs.~\eqref{gm2sfull} and \eqref{gm2s}. 
	Thus, $|C_{Ve} - C_{Ae} - C_{\nu e}| < 3.5 \times 10^{-4}$, while $|C_{V\mu} - C_{A\mu} - C_{\nu \mu}| > 4.6\times 10^{-4}$ in the mass range of interest, and hence there is no combination of effective couplings fulfilling Eq.~\eqref{cdifferenceFN} such that both anomalies are satisfied to within $1\sigma$ and all experimental constraints are satisfied. 
	It is notable that even in such a general theoretical setting, the $Z'$ explanation is unsuccessful.


	\section{$Z'$ solutions considering the Rubidium measurement}
	\label{sec:Rubidium}

	\begin{figure}
		\centering
		\includegraphics[width=0.5\textwidth]{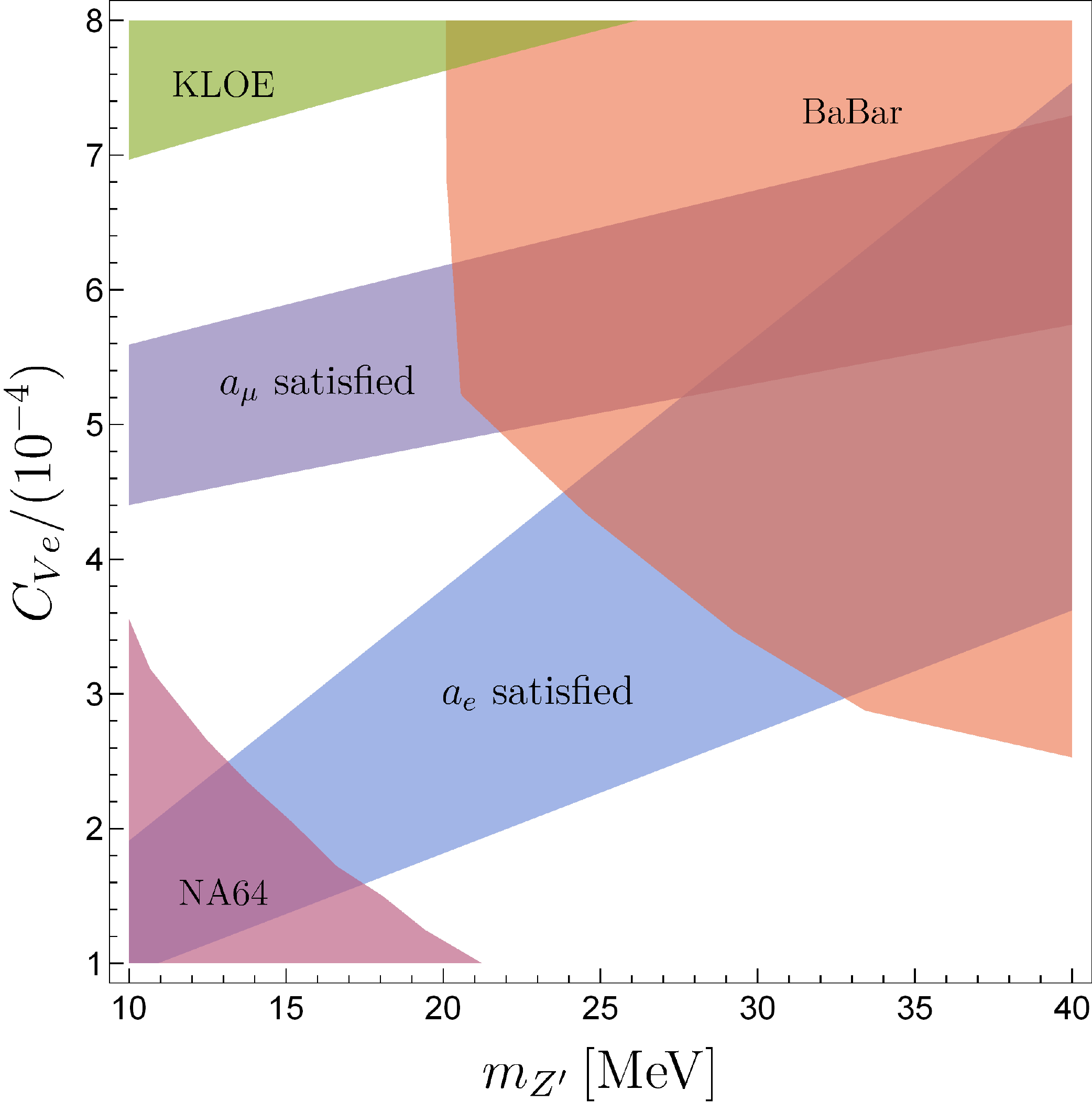}
		\caption{
			Constraints on the mass and effective coupling of the $Z'$, given the Rubidium measurement of $a_e$. 
			The bounds are on $C_{Ve}$, where we have taken $C_{Ae} = -10^{-8}/C_{Ae}$, saturating the M\o ller scattering constraint \eqref{moller}, and fixed $C_{V\mu}$ by Eq. \eqref{cdifferenceFN}, while setting all other effective couplings to zero. 
			The blue (purple) stripe corresponds to where $a_e$ ($a_\mu$) is satisfied to within $1\sigma$. 
			The pink region in the bottom left, green region in top left, and orange region in top right are ruled out by NA64, KLOE and BaBar, respectively. 
			As can be seen, the region simultaneously satisfying $a_e$ and $a_\mu$ is excluded by BaBar.}
		\label{fig:Rubid}
	\end{figure}

	We have thus far considered only the $(g-2)_e$ anomaly from the Caesium measurement, Eq. \eqref{electronanomaly}. 
	Significantly, this has the opposite sign to the muon anomaly. 
	In Section \ref{sec:models}, it was shown that the combination of the different signs and sizes of the anomalies, along with the copious experimental constraints, makes it impossible to construct a model which can satisfy both at the same time. 
	One might suppose that it is easier to explain two anomalies which have the same sign, which is exactly the situation if one considers instead the recent Rubidium result for $a_e$, cf. Eq. \eqref{electronanomalyRub}. 
	Here we consider this possibility. 
	This has not, to the best of our knowledge, previously been studied. 
	Our model-independent analysis of muon sector constraints in section \ref{muoncouplings} still applies. 
	The conclusions of the study of bounds on electron and electron neutrino couplings in section \ref{electroncouplings} are no longer valid, however Eqs.~\eqref{wdconstraint}-\eqref{borexino} all still hold.

	Let us immediately turn to the most general class of models considered in the previous section, the Froggatt-Nielsen scenario. 
	The SM$+U(1)'$ and NHDM$+U(1)'$ models are indeed specific cases of this set-up. 
	The key feature of this model is the relation between electron and muon couplings given in Eq.~\eqref{cdifferenceFN}, which is itself a consequence of gauge invariance. 
	We note that the magnitude of the Rubidium anomaly is similar to that of the Caesium anomaly, with $|\Delta a_e^\text{Rb}/\Delta a_e^\text{Cs}| = 0.55$, and therefore the former demands $C_{Ve} \sim \mathcal{O}(10^{-4})$, just as the latter had required $C_{Ae} \sim \mathcal{O}(10^{-4})$. 
	Moreover, the electron neutrino couplings are still constrained to be $\lesssim \mathcal{O}(10^{-5})$, with the bounds of Figs.~\ref{fig:e_constraints} (c,d) modified by an order-one factor because the relevant bounds are similar or identical under $C_{Ae} \to C_{Ve}$, see Eqs. \eqref{wdconstraint}, \eqref{texono} and \eqref{borexino}.

	The most minimal case is non-zero $C_{Ve}$ and $C_{V\mu}$ only, in which case Eq.~\eqref{cdifferenceFN} dictates $C_{Ve} = C_{V\mu}$ . Such a $Z'$ only satisfies both anomalies to within $1\sigma$ for $m_{Z'} \gtrsim 25$ MeV and $C_{Ve,\mu} \gtrsim 5 \times 10^{-4}$, which however is excluded by BaBar.\footnote{The BaBar and NA64 bounds on $|C_{Ae}|$ in Fig.~\ref{fig:e_constraints} (a,b) for $C_{Ve} \simeq 0$ (i.e. along the diagonal) can be reinterpreted here as a bound on $|C_{Ve}|$, since the experiments bound the combination $C_{Ve}^2 + C_{Ae}^2$.}  
	For smaller $m_{Z'}$, keeping the same effective coupling $C_V$ causes either too small a shift in $a_\mu$ or too great a shift in $a_e$.

	Generalising to include $C_{Ae}$ and $C_{A\mu}$, the former is restricted by the M\o ller scattering bound, $|C_{Ve} C_{Ae}| \lesssim 10^{-8}$. 
	In Fig.~\ref{fig:Rubid}, we plot 1$\sigma$ regions which explain the two anomalies individually along with the various constraints, setting $C_{Ae} = -10^{-8}/C_{Ve}$ to saturate the M\o ller scattering limit, and $C_{A\mu} = C_{\nu e} = C_{\nu \mu} = 0$. Eq.~\eqref{cdifferenceFN} dictates that $C_{V\mu} = C_{Ve} + 10^{-8}/C_{Ve}$. 
	As can be seen, while either anomaly can be satisfied by itself, the pair cannot simultaneously be explained. 
	Various alternatives do not ameliorate the problem. 
	Smaller $|C_{Ae}|$ would in turn require that $|C_{Ve}|$ is smaller in order to satisfy $\Delta a_e^\text{Rb}$, thereby lowering the blue bland in Fig.~\ref{fig:Rubid}. 
	Making $C_{Ae} > 0$ would decrease $C_{V\mu}$ as a function of $C_{Ve}$, thus raising the purple $a_\mu$ band. 
	Finally, larger $|C_{A\mu}|$ would mean a larger $C_{V\mu}$ is needed to explain $\Delta a_\mu$, this also raises the purple band. 
	For this reason, the general Froggatt-Nielsen scenario cannot solve the anomalies. 
	Since this set-up covers the SM$+U(1)'$ and NHDM$+U(1)'$ models, those scenarios are similarly unsuccessful.

	We see that three main challenges in explaining both $\Delta a_e^\text{Cs}$ and $\Delta a_\mu$\textemdash namely i) the relative magnitudes of the anomalies, ii) the stringent experimental limits on the different effective couplings, particularly $C_{\nu_e}$ and $C_{\nu_\mu}$, and iii) the relations between the effective couplings due to gauge invariance\textemdash are also present in the attempt to explain $\Delta a_e^\text{Rb}$ and $\Delta a_\mu$ simultaneously. 
	Thus, although the different signs of the muon and Caesium electron anomalies is an interesting feature, it therefore seems that this is not the main obstacle for $Z'$ model-building. 
	Since the sizes of the anomalies is fixed by experiment and the limits on effective couplings will only get stronger with time (see the summary in section \ref{sec:future}), in order to solve both anomalies one must find ways to get around Eq. \eqref{cdifferenceFN} in particular. 
	Possible ways to do this, such as introducing extra fermions, are beyond the scope of this paper.

	\section{Conclusion}
	There is a mixed experimental picture for the anomalous magnetic moment of charged leptons. 
	While the status of $(g-2)_\mu$ has been solidified by the recent Fermilab measurement, there is considerably more uncertainty surrounding $(g-2)_e$.
	We have explored in detail the possibility of simultaneously explaining both the (Caesium) $(g-2)_e$ and $(g-2)_\mu$ anomalies with a single low scale $Z'$. 
	After introducing the formalism in Section \ref{sec:Formalism}, in Section \ref{sec:constraints} we found the experimentally allowed region which can explain the anomalies to within $1\sigma$. 
	The permitted $Z'$ mass range is $16\textrm{ MeV} \lesssim m_{Z'} \lesssim 38 \textrm{ MeV}$, and one requires some sizeable effective couplings,
	$\{ 
	5\times 10^{-4} \lesssim |C_{V \mu }| \,\lesssim 0.05 ;~ 1.3 \times 10^{-4} \lesssim |C_{A e}| \lesssim 3.2 \times 10^{-4}
	\}$, and some smaller ones,
	$\{
	|C_{V e}| \lesssim 7.7 \times 10^{-5};~
	|C_{\nu e}, C_{\nu\mu}| \lesssim 10^{-5}
	\}$,
	while $|C_{A \mu }|$ can be anywhere between 0 and $8 \times 10^{-3}$ depending on the size of $|C_{V\mu}|$.
	The key findings are summarised in Figs.~\ref{fig:e_constraints} and \ref{fig:mu_constraints}. 
	Our survey of the parameter space was very general, in particular allowing for both vector and axial $Z'$ couplings and for flavour non-universality. 
	Turning to the range of experiments planned for the near future, we argued in Section \ref{sec:future} that the entirety of the allowed parameter space for solving the $(g-2)_e$ anomaly could be tested soon, in particular by NA64, Belle-II and MAGIX.

	This analysis provides a very specific target for model-building. 
	In Section \ref{sec:models} we explored three classes of models of increasing complexity with the aim of generating a combination of couplings which lies within the allowed parameter space. 
	In the simplest extension, a SM+$U(1)'$ model, the gauge invariance of the Yukawa couplings prevented the significant $C_{Ae}$ required to resolve the $(g-2)_e$ anomaly. 
	Going further to a 2HDM+$U(1)'$ (which can be generalised to a NHDM+$U(1)'$ scenario), we showed that the smallness of the neutrino couplings required to evade constraints from neutrino scattering experiments demands nearly universal effective vector couplings, and that this, in addition to the universal effective axial couplings of the model, does not permit an explanation of both the anomalies at the same time. 
	Finally, we turned to a Froggatt-Nielsen inspired scenario which permitted greater freedom by removing the gauge invariance of the Yukawas. 
	The relation between the couplings to left-handed charged leptons and their respective neutrinos imposed by the gauge structure, in conjunction with the very stringent bounds on the neutrino couplings in particular, again conspired to forbid a solution to the two anomalies.

	We then demonstrated in Section \ref{sec:Rubidium} that such models also cannot simultaneously satisfy the $(g-2)_\mu$ and Rubidium $(g-2)_e$ anomalies. 
	This was notable since those two anomalies have the same sign. 
	Thus, factors such as the strong individual limits on $Z'$ couplings (studied in Section \ref{sec:constraints}) and the relative size of the two anomalies are more challenging to overcome in $Z'$ models than their relative sign. 
	To our knowledge, this was the first study of a $Z'$ explanation for the muon anomaly with the newest $(g-2)_e$ result. 
	The conclusion of our analysis is that $Z'$-only explanations of the dual $(g-2)_e$ and $(g-2)_\mu$ anomalies are ruled out. 
	Additional new fields must be introduced in order to explain the two discrepancies. 
	This is true both for the Caesium and Rubidium values of $a_e$. 
	

	If the $(g-2)_\mu$ anomaly, measured both at Brookhaven and Fermilab, is borne out by the future J-PARC experiment, and (either) $(g-2)_e$ discrepancy persists, the SM will be faced by two disagreements between theory and experiment of a similar nature but a different magnitude and possibly sign. 
	In principle, a MeV-scale vector boson can have couplings to leptons which resolve both while satisfying the plethora of existing experimental constraints. 
	It appears, however, that additional fields contributing to leptonic magnetic moment(s) are also required. 
	Given the promising experimental outlook over the next decade, we should know soon whether or not there does exist such a $Z'$, and associated dark sector, with the ability to resolve the $\anomalies$ anomalies.
	\\

	
	\noindent \textbf{Acknowledgements}\\
	\\
	We thank Stefano Moretti and Raman Sundrum for their very helpful comments on the manuscript. A.B. and R.C. thank the organisers of the 2019 SLAC Summer Institute, `Menu of Flavors', where this project was conceived and initiated. In particular, we thank Thomas G. Rizzo for the encouragement to pursue this work, and also Felix Kress, Elisabeth Niel, Peilong Wang and Jennifer Rittenhouse West for interesting discussions. 
	A.B. is supported by the NSF grant PHY-1914731 and by the Maryland Center for Fundamental Physics (MCFP). R.C. is supported by the IISN convention 4.4503.15. 
	R.C. thanks the UNSW School of Physics, where he is a Visiting Fellow, for their hospitality during part of this project.
	
	\bibliographystyle{JHEP}
	\bibliography{gm2_refs}
	
\end{document}